\documentclass[
amsmath,
amssymb,
aps,
twocolumn,
floatfix,
superscriptaddress,
longbibliography,
]{revtex4-1}

\usepackage[dvipdfmx]{graphicx}
\usepackage{dcolumn}
\usepackage{bm}
\usepackage{braket}
\usepackage{amsthm}
\usepackage{natbib}
\usepackage{xcolor}
\usepackage{listings}
\DeclareMathOperator*{\argmax}{arg\,max}

\usepackage[
colorlinks=true,
urlcolor=blue,
citecolor=blue,
linkcolor=blue,
hyperfootnotes=false,
pdfencoding=auto
]{hyperref}

\begin{document}

\title{
Diamond-shaped quantum circuit for real-time quantum dynamics in one dimension
}

% Authors

\author{Shohei Miyakoshi}
\email{miyakoshi.shohei.qiqb@osaka-u.ac.jp}
\affiliation{Center for Quantum Information and Quantum Biology, Osaka University, Toyonaka, Osaka, 560-8531, Japan}
\affiliation{Computational Materials Science Research Team, RIKEN Center for Computational Science (R-CCS), Kobe, Hyogo, 650-0047, Japan}

\author{Takanori Sugimoto}
\affiliation{Center for Quantum Information and Quantum Biology, Osaka University, Toyonaka, Osaka, 560-8531, Japan}
\affiliation{Computational Materials Science Research Team, RIKEN Center for Computational Science (R-CCS), Kobe, Hyogo, 650-0047, Japan}
\affiliation{Advanced Science Research Center, Japan Atomic Energy Agency, Tokai, Ibaraki 319-1195, Japan}

\author{Tomonori Shirakawa}
\affiliation{Computational Materials Science Research Team, RIKEN Center for Computational Science (R-CCS), Kobe, Hyogo, 650-0047, Japan}
\affiliation{Quantum Computational Science Research Team, RIKEN Center for Quantum Computing (RQC), Wako, Saitama, 351-0198, Japan}
\affiliation{Computational Condensed Matter Physics Laboratory, RIKEN Cluster for Pioneering Research (CPR), Wako, Saitama, 351-0198, Japan}
\affiliation{RIKEN Interdisciplinary Theoretical and Mathematical Science Program (iTHEMS), Wako, Saitama, 351-0198, Japan}

\author{Seiji Yunoki}
\affiliation{Computational Materials Science Research Team, RIKEN Center for Computational Science (R-CCS), Kobe, Hyogo, 650-0047, Japan}
\affiliation{Quantum Computational Science Research Team, RIKEN Center for Quantum Computing (RQC), Wako, Saitama, 351-0198, Japan}
\affiliation{Computational Condensed Matter Physics Laboratory, RIKEN Cluster for Pioneering Research (CPR), Wako, Saitama, 351-0198, Japan}
\affiliation{Computational Quantum Matter Research Team, RIKEN Center for Emergent Matter Science (CEMS), Wako, Saitama 351-0198, Japan}

\author{Hiroshi Ueda}
\affiliation{Center for Quantum Information and Quantum Biology, Osaka University, Toyonaka, Osaka, 560-8531, Japan}
\affiliation{Computational Materials Science Research Team, RIKEN Center for Computational Science (R-CCS), Kobe, Hyogo, 650-0047, Japan}

\begin{abstract}
In recent years, quantum computing has evolved as an exciting frontier, with the development of numerous algorithms dedicated to constructing quantum circuits that adeptly represent quantum many-body states.
However, this domain remains in its early stages and requires further refinement to understand better the effective construction of highly-entangled quantum states within quantum circuits.
Here, we demonstrate that quantum many-body states can be universally represented using a quantum circuit comprising 
multi-qubit gates.
Furthermore, we evaluate the efficiency of a quantum circuit constructed with two-qubit gates in quench dynamics for the transverse-field Ising model. 
In this specific model, despite the initial state being classical without entanglement, it undergoes long-time evolution, 
eventually leading to a highly-entangled quantum state.
Our results reveal that a diamond-shaped quantum circuit, designed to approximate the multi-qubit gate-based quantum circuit, remarkably excels in accurately representing the long-time dynamics of the system. 
Moreover, the diamond-shaped circuit follows the volume law behavior in entanglement entropy, offering a significant advantage over alternative quantum circuit constructions employing two-qubit gates. 
\end{abstract}


\date{\today}

\maketitle
%\tableofcontents

\section{\label{sec:intro} Introduction}

%%% Algorithms for FTQC

The recent strides made in state-of-the-art quantum computers have ushered in a new era of possibilities for tackling a diverse array of complex problems in computational science.
In particular, the progress towards achieving long-term fault-tolerant quantum computers (FTQC) has intensified interest in designing efficient quantum circuits for general-purpose computing across various disciplines, including quantum chemistry~\cite{peruzzo_variational_2014,yung_transistor_2014,omalley_Scalable_2016,kandala_Hardwareefficient_2017,romero_Strategies_2018,evangelista_Exact_2019,mcardle_Quantum_2020} and quantum machine learning~\cite{biamonte_Quantum_2017,mitarai_Quantum_2018,perdomo-ortiz_Opportunities_2018}. 
Nevertheless, optimizing the structure of quantum circuits, which involves managing numerous internal parameters to represent highly-entangled quantum states, continues to pose a significant 
challenge~\cite{grimsley_adaptive_2019,tang_QubitADAPTVQE_2021,liu_efficient_2021,yao_Adaptive_2021,zhang_Adaptive_2021,shirakawa_Automatic_2021,lin_Real_2021,haghshenas_Variational_2022}.

%%% Additional difficulty in NISQ

Within quantum computing, noisy intermediate-scale quantum (NISQ) devices currently face significant challenges due to severe limitations in quantum-circuit depth, the number of qubits, and the number of gate operations, all caused by inevitable quantum noise~\cite{arute_Quantum_2019,zhou_What_2020,franca_Limitations_2021}.
Therefore, executing quantum-circuit algorithms on NISQ devices presents another challenge: implementing entangled states in {\it shallow} circuits with a certain degree of accuracy. 
This challenge is particularly pertinent when considering computational resources, such as computation time and storage, even in the long-term FTQCs. 
Most contemporary research utilizing NISQ devices reports results for a low-entangled quantum state or a matrix-product state (MPS) with a small bond dimension represented by a shallow quantum circuit~\cite{smith_Crossing_2022}.
However, to achieve scalable quantum computing, overcoming this challenge is imperative.

%%% importance of the efficient shape of the quantum circuit 

For the successful realization of highly-entangled quantum states in NISQ devices, the quantum circuits representing these states must consist of a polynomial number of gate operations, with an expressivity comparable to MPS with a large number of bond dimensions, while ensuring a defined level of precision.
Notably, stacking multiple gate operations may increase the effective bond dimensions in terms of MPS; however, this approach potentially complicates the optimization of internal parameters and the management of these circuits within NISQ devices.
Therefore, it is essential to identify the efficient network of quantum circuits and the complexity class of highly-entangled quantum states that shallow quantum circuits cannot effectively represent. 
Such understandings deepen our knowledge of quantum many-body states in condensed matter physics and non-equilibrium quantum dynamics, significantly contributing to advancements in quantum computing.

%%$ this study$

In this study, we present an analysis of global quench dynamics for the quantum Ising model in both transverse and longitudinal fields, employing various types of quantum-circuit states. 
Recent literature on this study has explored similar models using different quantum devices, such as trapped atomic ions~\cite{zhang_Observation_2017} and superconducting qubits~\cite{kim_Evidence_2023}. 
Despite these experiments facing limitations in the number of gate operations due to noise accumulation, they demonstrate the feasibility of evaluating physical quantities of highly-entangled quantum states, which may pose challenges for current classical computations~\cite{Tindall_2023,Kechedzhi_2023,begusic_Fast_2023,Liao_2023,Begusic_2023,Patra_2023}.
In contrast to these investigations, our approach focuses on achieving long-time evolution by compressing 
a time-evolved quantum state into a quantum-circuit representation with a reduced gate count. 

Our investigation begins by examining quantum circuits composed of multi-qubit gates.  
Specifically, we assess the gate size of multi-qubit gates included in a quantum circuit, 
which is essential for accurately representing a quantum state in non-equilibrium dynamics.
Our numerical analyses suggest that a quantum circuit employing multi-qubit gates greater than half the system size 
can accurately represent the numerically exact long-time dynamics. 
However, the use of large multi-qubit gates poses optimization challenges due to the exponential increase of internal parameters 
with increasing the system size. 
To address this issue, we explore the possibility of decomposing a multi-qubit gate into two-qubit gates in two distinct quantum circuit configurations: sequential-type and diamond-shaped quantum circuits.
Through an exhaustive numerical assessment of these configurations, we reveal that the diamond-shaped quantum circuit exhibits superior fidelity in describing the non-equilibrium quantum dynamics. 
This result is noteworthy as the diamond-shaped quantum circuit requires fewer gates than its sequential-type counterpart. 

The rest of this paper is organized as follows. 
In Sec.~\ref{sec:model}, we introduce a model for non-equilibrium quantum dynamics and elaborate on the optimization schemes of a quantum circuit to describe the real-time evolution of the system.
In Sec.~\ref{sec:multiqubit}, we demonstrate the real-time evolution using a quantum circuit with multi-qubit gates for the global quench dynamics. 
In Sec.~\ref{sec:diamond}, we introduce a diamond-shaped quantum circuit and examine its fidelity to the numerically exact state.  
We also compare the results with other types of quantum circuits. 
Finally, we summarize the results and give insights from this study in Sec.~\ref{sec:conclusion}. 
Additionally, we summarize a quantum circuit optimization scheme designed for a time-evolved state in Appendix~\ref{app:method}. 
Appendix~\ref{app:exact-circuit} proves that a quantum circuit with multi-qubit gates can represent any quantum state.  
We discuss the upper bound of entanglement entropy for a quantum-circuit state in Appendix~\ref{app:entanglement}. 
Furthermore, we provide more numerical results for the sequential-type quantum circuits in Appendix~\ref{app:optimize_each_timestep}.

\section{\label{sec:model} Model and method}

Our research explores various types of quantum-circuit ansatze, with a special focus on their ability to accurately represent quantum dynamics.
To achieve this, we make a comprehensive analysis of the global quench dynamics for a quantum Ising model.
This model includes both the transverse and longitudinal fields, controlled by the dimensionless variables $g$ and $h$, respectively, as described in the following Hamiltonian:
\begin{equation}
\hat{H} = \sum_{j=1}^{L-1} \hat{h}_{j,j+1} + \hat{h}_\mathrm{left} + \hat{h}_\mathrm{right}
\end{equation}
with
\begin{align}
& \hat{h}_{j,j+1} = -J\left[\hat{\sigma}^{x}_{j}\hat{\sigma}^{x}_{j+1}
+ \sum_{k=0,1}\left( \frac{g}{2}\hat{\sigma}^{z}_{j+k} + \frac{h}{2}\hat{\sigma}^{x}_{j+k} \right) \right] 
\end{align}
and
\begin{align}
& \hat{h}_\mathrm{left} = -J \left( \frac{g}{2} \hat{\sigma}^{z}_{1} + \frac{h}{2} \hat{\sigma}^{x}_{1} \right) , \\
& \hat{h}_\mathrm{right} = -J \left( \frac{g}{2} \hat{\sigma}^{z}_{L} + \frac{h}{2} \hat{\sigma}^{x}_{L} \right) , 
\end{align}
where the Ising coupling $J$ is positive and $\hat{\sigma}^{\alpha}_{j}$ ($\alpha=x,z$) is the $\alpha$-component 
of the Pauli operator at site $j$ on a one-dimensional (1D) chain of $L$ sites. 
We assume that the initial state at time $t=0$ is given by the spin-polarized product state
$\ket{\Phi}=\ket{\uparrow\uparrow\cdots\uparrow}$.
When $h=0$, the model is equivalent to the transverse-field Ising model, which is integrable.
As the longitudinal field $h$ increases, the quantum dynamics after the global quench becomes more chaotic, and the fast scrambling 
makes the classical simulation difficult~\cite{lin_Real_2021}.

The primary focus of our research is the real-time dynamics of the quantum state $\ket{\Psi(t)}=e^{-i\hat{H} t} \ket{\Phi}$. 
In order to implement the time evolution on a quantum computer, we utilize a second-order Trotter decomposition of 
$e^{-i\hat{H} \Delta t}$ with a time step $J\Delta t \ll 1$, i.e., 
\begin{align}
&\hat{V}(\Delta t)=e^{-i\hat{H}_\mathrm{odd}\Delta t/2}e^{-i\hat{H}_\mathrm{even}\Delta t}
e^{-i\hat{H}_\mathrm{odd}\Delta t/2}, 
\end{align}
where $\hat{H}=\hat{H}_\mathrm{even}+\hat{H}_\mathrm{odd}$ with 
\begin{align}
& \hat{H}_\mathrm{even} = \left\{ \begin{matrix}\displaystyle 
\sum^{L/2-1}_{j=1} \hat{h}_{2j,2j+1} & (L:\,\mathrm{even}) \\ 
\displaystyle
\sum^{(L-1)/2}_{j=1} \hat{h}_{2j,2j+1} + \hat{h}_\mathrm{left} & (L:\,\mathrm{odd}) \\ 
\end{matrix} \right. 
\end{align}
and 
\begin{align}
& \hat{H}_\mathrm{odd} = \left\{ \begin{matrix} \displaystyle
\sum^{L/2}_{j=1} \hat{h}_{2j-1,2j} + \hat{h}_\mathrm{left} + \hat{h}_\mathrm{right} & (L:\,\mathrm{even}) \\ 
\displaystyle
\sum^{(L-1)/2}_{j=1} \hat{h}_{2j-1,2j} + \hat{h}_\mathrm{right} & (L:\,\mathrm{odd}) \\ 
\end{matrix} \right. .
\end{align}
% Notice that each operator in $\hat{H}_\mathrm{even}$ ($\hat{H}_\mathrm{odd}$) commutes with other operators. 
The operator $\hat{V}(\Delta t)$ is then applied successively to $\ket{\Phi}$ to obtain the quantum state at time $t$, 
i.e., $\ket{\Psi(t)} = \hat{V}(\Delta t)\ket{\Psi(t-\Delta t)} + O(\Delta t^3)$ with $|\Psi(t=0)\rangle = |\Phi\rangle$. 
Such a quantum dynamics process poses a canonical problem in attaining quantum acceleration~\cite{zhang_Observation_2017,kim_Evidence_2023}.
This is because, as the circuit depth increases along the temporal axis, the bipartite entanglement entropy generally rises monotonically, eventually reaching a value proportional to the volume of the subsystem.
This phenomenon is widely recognized as the volume law.   
On the other hand, most classical approaches are sophisticated to treat quantum states satisfying the entropic area law, i.e., the entanglement entropy being proportional to the boundary area of the subsystem, as opposed to its volume.
Therefore, it is challenging to achieve long-time dynamics using classical approaches.

\begin{figure}[htbp]
\includegraphics[width=0.75\columnwidth]{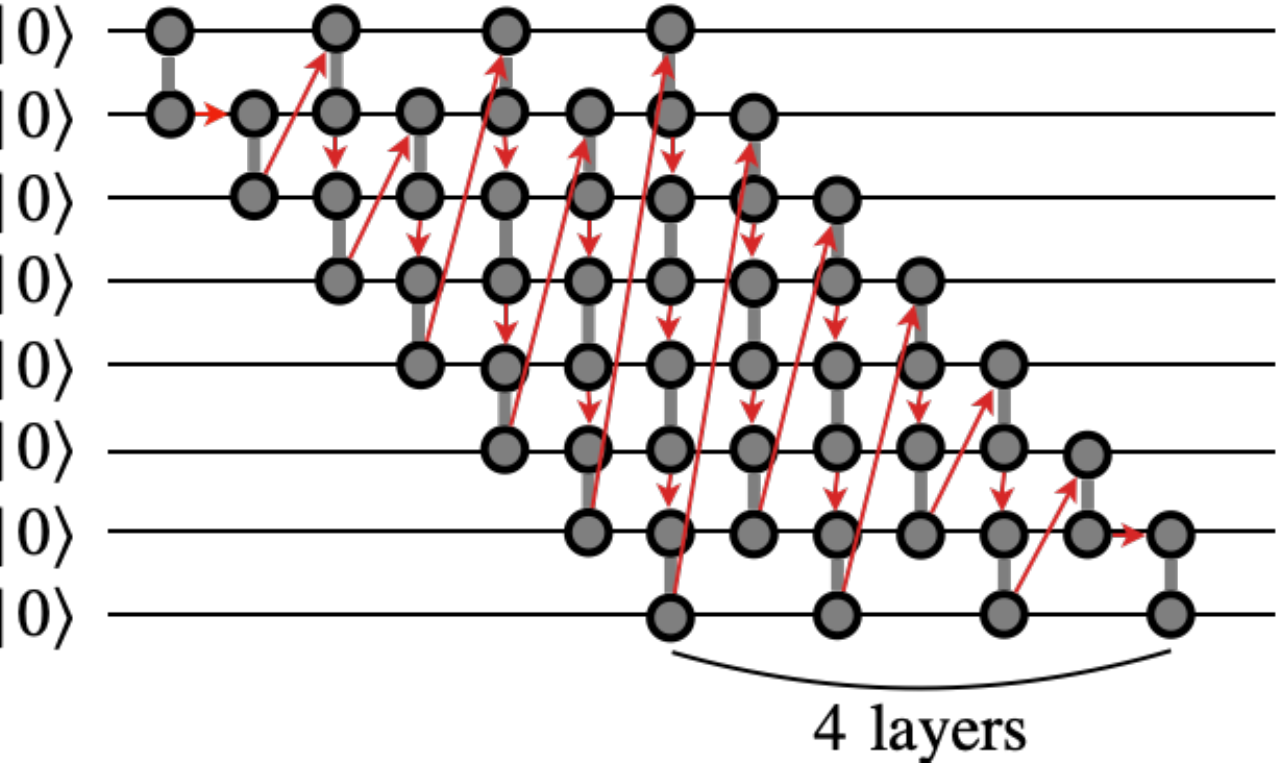}
\caption{
An example of the quantum-circuit ansatz: a sequential-type quantum circuit with $4$ layers for $L=8$.
Each gray dumbbell-shaped symbol represents a two-qubit gate corresponding to a U(4) operator, and the red arrows between two-qubit gates indicate the optimization sequence for these gates in our numerical calculations.
}
\label{fig:circuit-sequential-order}
\end{figure}

\begin{figure}[htbp]
\includegraphics[width=\columnwidth]{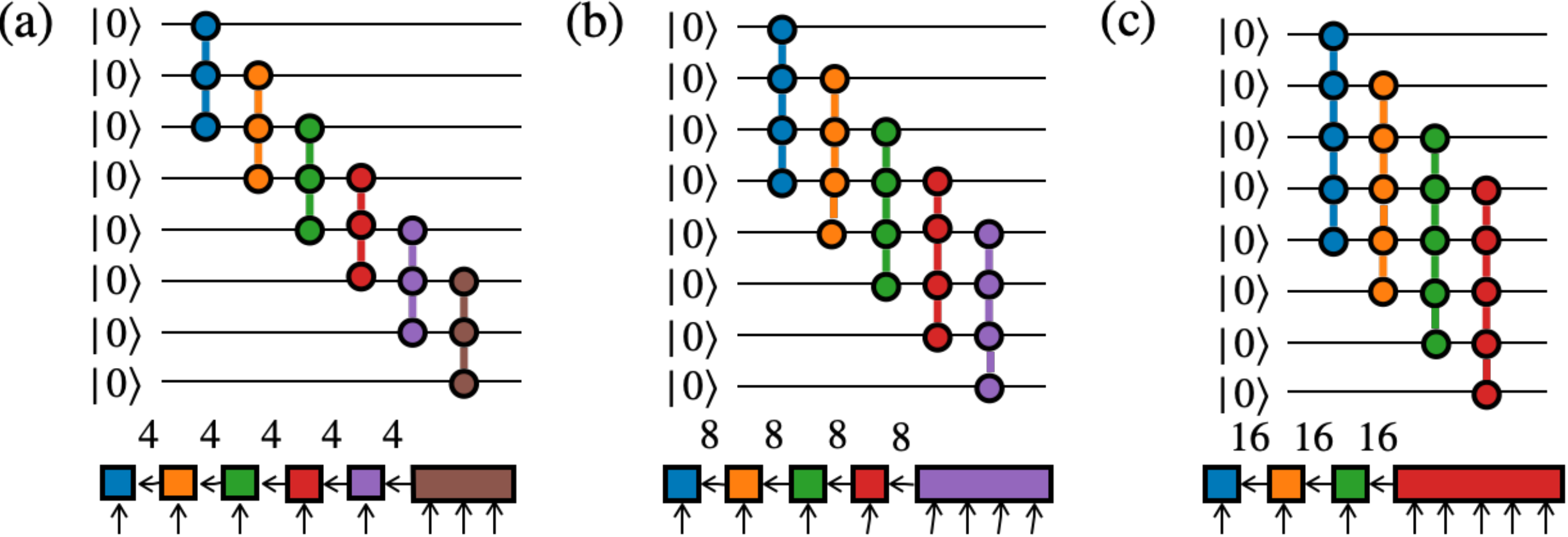}
\caption{
The single-layer sequential-type quantum circuits consisting of (a) three-qubit gates ($l=3$), (b) four-qubit gates ($l=4$), 
and (c) five-qubit gates ($l=5$) for $L=8$. 
The tensor-network representations for these quantum circuits are also shown in the bottom panels, where 
each colored square or rectangle corresponds to the multi-qubit gate with the same color in the quantum circuit, and 
each virtual bond in the tensor-network state represents the internal degrees of freedom for qubits connecting the 
neighboring multi-qubit gates. 
The number above each bond represents the bond dimension. 
The arrows in the tensor-network state indicate the direction from the local physical states to the top tensor (indicated by a blue square).
It is straightforward to extend these quantum circuits with the corresponding tensor-network representations 
to those composed of different $l$-qubit gates for other system sizes.
For example, the quantum circuit depicted in Fig.~\ref{fig:circuit-sequential-order} is a four-layer sequential-type quantum circuit 
composed of two-qubit gates for $L=8$.  
}
\label{fig:circuit-multi-qubit}
\end{figure}

For NISQ devices, dealing with excessively deep quantum circuits presents challenges due to the inherent noise and decoherence, which hampers the direct implementation of the long-time evolution of quantum many-body dynamics.
Hence, a variational approach using a parametrized quantum circuit $\hat{C}_{t}$ with constant depth emerges as a viable strategy. 
As a case in point, the sequential-type quantum circuit often adopted for analyzing 1D quantum many-body systems is shown in Fig.~\ref{fig:circuit-sequential-order}.
Here, we consider a quantum circuit $\hat{C}_{t}$ consisting of a product of U(4) operators, i.e., 2-qubit gates. 
To further elaborate on the expressivity of parametrized quantum circuits in the context of quantum dynamics, we expand the gate size to act on $l=3,4,5$ qubits in the sequential-type quantum circuits, as shown schematically in Fig.~\ref{fig:circuit-multi-qubit} (see Sec.~\ref{sec:multiqubit} for the detailed explanation).
Additionally, in Sec.~\ref{sec:diamond}, we introduce a diamond-shaped quantum circuit as the optimal configuration for describing quantum many-body dynamics.

Employing a variety of quantum circuit configurations, we approximately represent the real-time evolved quantum state $|\Psi(t)\rangle \approx \hat{C}_{t} |\Psi\rangle$ by iteratively optimizing the quantum circuit $\hat{C}_{t}$ to maximize the fidelity,
\begin{align}
&I(t)=\left|\bra{\Phi}\hat{C}_{t-\Delta t}^\dag\hat{V}(\Delta t)^{\dagger}\hat{C}_{t}\ket{\Phi}\right|, 
\label{eq:i}
\end{align}
where $\hat{C}_{t-\Delta t}$ is already optimized in the previous time step for approximately representing 
$|\Psi(t-\Delta t)\rangle$. 
To address this variational problem, we sequentially optimize each gate one by one in the quantum circuit, as often employed in tensor network methods.
More details of the optimization technique can be found in Appendix~\ref{app:method}.
The optimization sequence is outlined in the following.
As illustrated in Fig.~\ref{fig:circuit-sequential-order}, the optimization begins with the gate in the top-left corner.
Subsequently, the gates are optimized sequentially, moving from left to right and from top to bottom, especially when there are multiple vertically stacked gates.
This sequence continues until one reaches the gate in the bottom-right corner. 
Upon completing this step, the optimization direction reverses, iterating through the gates in the inverse order 
until returning to the initial gate in the top-left corner.
This entire process constitutes a single round of optimization, termed a ``single sweep update", hereafter.
This process is repeated multiple times (i.e., through many sweep updates) until the convergence is achieved.

For the termination criteria of the optimization, we introduce four hyperparameters: the upper (lower) limit $w_\mathrm{max}$ ($w_\mathrm{min}$) on the number of sweep updates, and the absolute (relative) convergence threshold $\epsilon_{a}$ ($\epsilon_{r}$).
Utilizing these hyperparameters, the termination condition is defined as $|1-I(t)|<\epsilon_{a}$ and $|1-I(t)/I(t-\Delta t)|<\epsilon_{r}$, within the range of sweep updates $[w_\mathrm{min},w_\mathrm{max}]$. 
Therefore, the sweep update is executed a minimum of $w_\mathrm{min}$ times while it is capped at a maximum of 
$w_\mathrm{max}$ times. 
For the optimization of the quantum circuit $\hat{C}_t$ at time $t$ in Eq.~(\ref{eq:i}), 
we use the quantum circuit $\hat{C}_{t-\Delta t}$, which has already been optimized during the previous time step, 
as the initial circuit for the optimization iteration.
This applies for all time steps except when $t=0$, where the identity operator is assigned to each qubit gate as the initial condition. 
The time step is set at $\Delta t=0.01/J$, chosen to be sufficiently small to ensure that the Trotter error has negligible effect on the numerical accuracy of the results presented below.

\section{\label{sec:multiqubit}Multi-qubit quantum circuit}

We begin by evaluating the expressivity of a single-layer sequential-type quantum circuit consisting of multi-qubit gates (see top panels in Fig.~\ref{fig:circuit-multi-qubit}) in representing real-time quantum dynamics. 
Here, the size of a multi-qubit gate is denoted as $l$, and thus each gate corresponds to a U($2^l$) operator.
Expanding the size of these gates enhances the capability to represent complex quantum states, provided that the system size $L$ remains finite~\cite{miyamoto_Extracting_2022}.
However, in practice, optimizing these quantum circuits frequently encounters the barren-plateau problem~\cite{mcclean_Barren_2018}.

\begin{figure}[htbp]
\includegraphics[width=1.0\columnwidth]{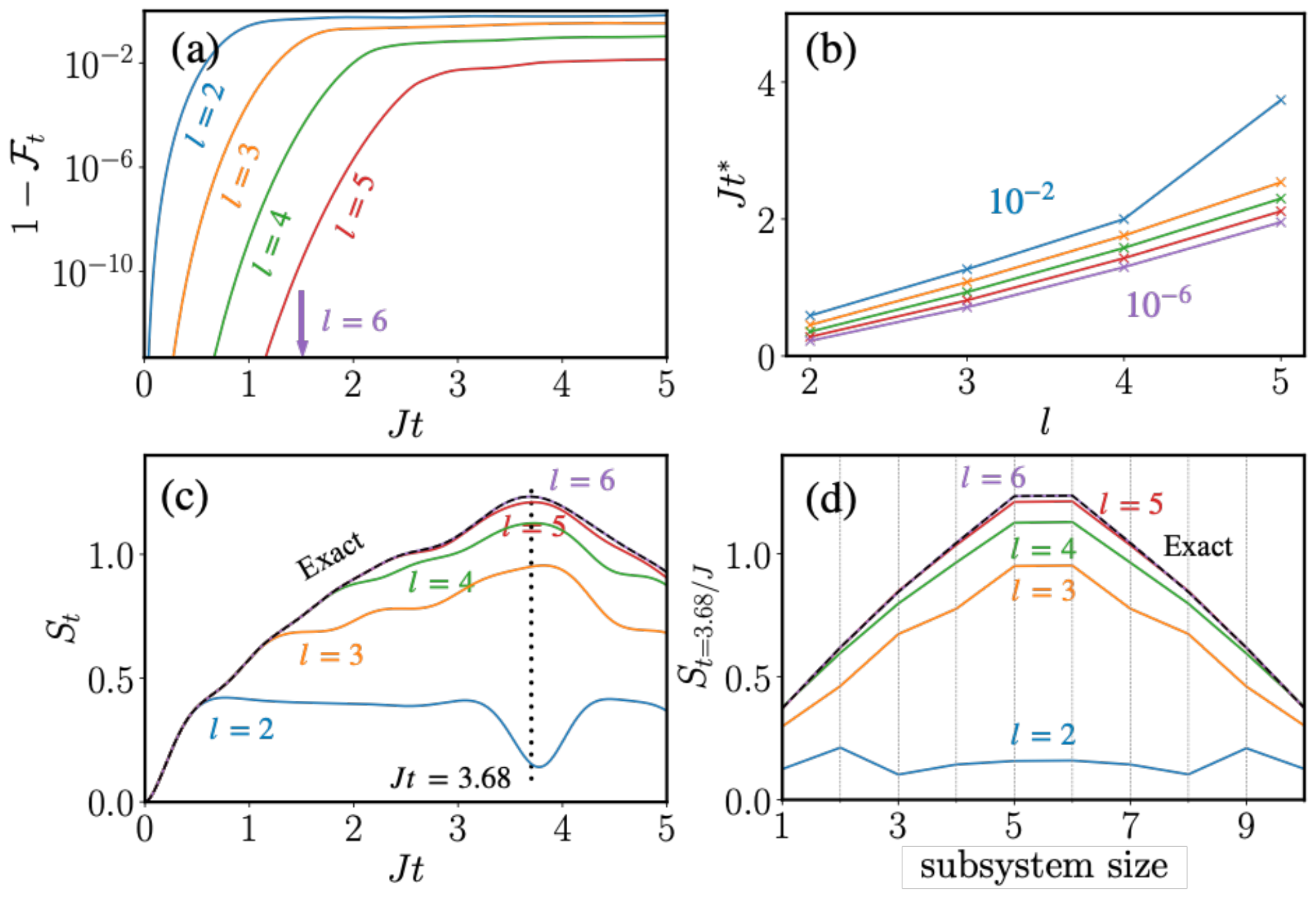}
\caption{
The infidelity and von Neumann entropy in the global quench dynamics for the $L=11$ quantum Ising chain with 
the model parameters $g=1.4$ and $h=0.1$ in the non-integrable region. The time-evolved state $|\Psi(t)\rangle$ is approximated 
using the single-layer sequential-type quantum circuits consisting of $l$-qubit gates, $|\Phi(t)\rangle=\hat{C}_t|\Phi\rangle$ 
(see Fig.~\ref{fig:circuit-multi-qubit}). 
(a) Time evolution of the infidelity $1-\mathcal{F}_t$ for different gate sizes.
The arrow indicates the infidelity $1-\mathcal{F}_t$ lower than $10^{-14}$.
(b) The maximum evolution time $t^{*}$ below which $1-\mathcal{F}_t<10^{-z}$ ($z=2,3,\cdots,6$) as a function of the gate size $l$.
(c) Time evolution of the von Neumann entropy $S_t$ for different gate sizes. Here, the system is partitioned into subsystems of 5 and 6 sites.
The vertical dotted line highlights the evolution time at which the numerically exact time-evolved state $|\Psi(t)\rangle$ 
exhibits the maximal von Neumann entropy, i.e., $Jt=3.68$. 
(d) The subsystem size dependence of the von Neumann entropy $S_t$ evaluated at $Jt=3.68$. 
For comparison, the numerically exact results evaluated from $|\Psi(t)\rangle$ are also plotted by the dashed lines in (c) and (d). 
The hyperparameters are set to be $w_\mathrm{max}=10^{4}$, $w_\mathrm{min}=10^{2}$, $\epsilon_{a}=10^{-14}$, and $\epsilon_{r}=10^{-4}$. 
}
\label{fig:compare-multi-qubit-l11-l16}
\end{figure}

\begin{figure}[htbp]
\includegraphics[width=1.0\columnwidth]{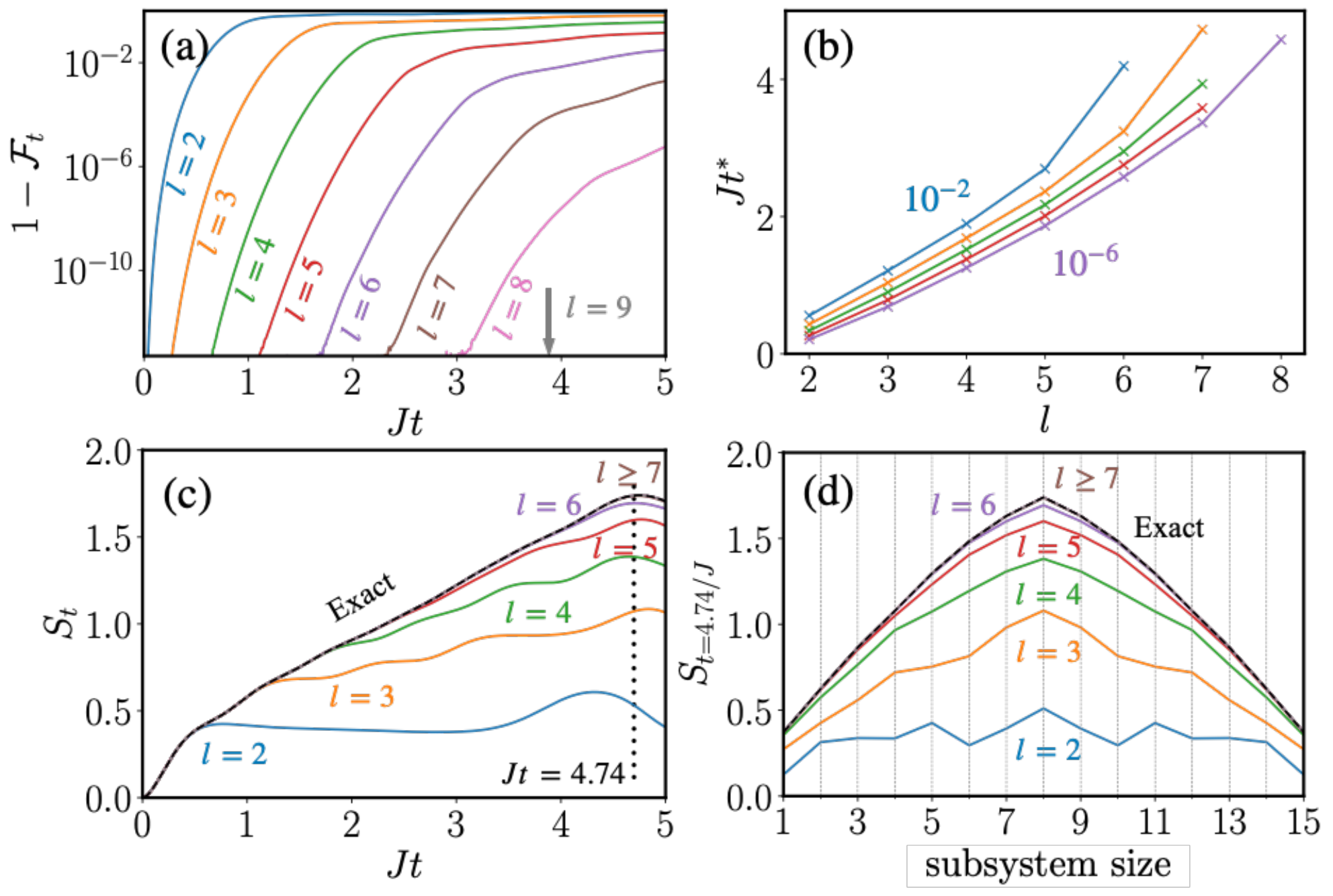}
\caption{
Same as Fig.~\ref{fig:compare-multi-qubit-l11-l16} but for $L=16$. 
$S_t$ in (c) is evaluated by bipartitioning the system exactly half. 
The vertical dot line in (c) indicates $Jt=4.74$, and $S_t$ in (d) is evaluated at $Jt=4.74$. 
}
\label{fig:compare-multi-qubit-l11-l16_2}
\end{figure}

Figures~\ref{fig:compare-multi-qubit-l11-l16}(a) and \ref{fig:compare-multi-qubit-l11-l16_2}(a) show the $l$ dependence 
of the infidelity $1-\mathcal{F}_t$ for $L=11$ and $16$, respectively, where
\begin{align}
\mathcal{F}_t=\left|\langle\Psi(t)|\Phi(t)\rangle\right|^2 
\end{align}
with $|\Psi(t)\rangle$ and $\ket{\Phi(t)}$ being the exact and approximate time-evolved quantum states given by 
$|\Psi(t)\rangle=e^{-i\hat{H}t}|\Phi\rangle$ and $\ket{\Phi(t)}=\hat{C}_t\ket{\Phi}$. 
In this study, the model parameters were set to $g=1.4$ and $h=0.1$ in the non-integrable region.
As observed in these figures, the infidelity increases monotonically with time $t$, irrespective of $l$, except for $l=6$ ($l=9$) in the 
$L=11$ ($L=16$) system (see below for more discussion), but it decreases with increasing $l$ for all time $t$. 

Furthermore, we estimate the maximum evolution time $t^{*}$ below which the infidelity satisfies $1-\mathcal{F}_t<10^{-z}$ $(z=2,3,\cdots,6)$.
Figures~\ref{fig:compare-multi-qubit-l11-l16}(b) and \ref{fig:compare-multi-qubit-l11-l16_2}(b) show the maximum evolution time $t^*$ as a function of the gate size $l$ for $L=11$ and $16$, respectively.
We observe that the maximum evolution time $t^*$ increases almost linearly with increasing the gate size $l$. 
Since the fidelity $\mathcal{F}_t$ is an index of the expressivity of quantum circuits, these results numerically confirm that quantum circuits improve their expressivity by increasing the gate size $l$.

It is also interesting to notice in 
Figs.~~\ref{fig:compare-multi-qubit-l11-l16}(a) and \ref{fig:compare-multi-qubit-l11-l16_2}(a) that when the gate size $l$ is larger than half the system size, i.e., $l=6$ for $L=11$ and $l=9$ for $L=16$, the quantum circuit can reproduce numerically exact quantum dynamics in any time region. 
This can be explained through the tensor-network representation of a quantum circuit.
Figure~\ref{fig:circuit-multi-qubit} shows the single-layer sequential-type quantum circuit composed of different multi-qubit gates with $l=3,4$ and 5 for $L=8$, along with the corresponding tensor-network representations. 
As elaborated in Appendix~\ref{app:exact-circuit}, the single-layer sequential-type quantum circuit composed of $l$-qubit gates is equivalent to the right-canonical matrix-product state (MPS) with a fixed bond dimension $\chi=2^{l-1}$. 
Therefore, when $l=\lfloor \frac{L}{2}\rfloor+1$, the bond dimension of the corresponding MPS is $\chi=2^{\lfloor \frac{L}{2}\rfloor}$ and 
hence, the MPS with this bond dimension can represent any quantum state of the system with $L$ sites.

We now investigate the gate-size dependence of the von Neumann entropy $S_t=-\mathrm{Tr}_B[\hat{\rho}_B(t)\ln\hat{\rho}_B(t)]$ with the reduced density matrix $\hat{\rho}_B(t) = \mathrm{Tr}_{\bar{B}}\left[ \ket{\Phi(t)} \bra{\Phi(t)} \right]$, where $\mathrm{Tr}_{B(\bar{B})}$ denotes the trace of the degrees of freedom associated with subsystem $B$ (the complement of subsystem $B$) partitioning the whole 1D system. 
Figures~\ref{fig:compare-multi-qubit-l11-l16}(c) and \ref{fig:compare-multi-qubit-l11-l16_2}(c) show the gate-size dependence of the von Neumann entropy $S_t$ for $L=11$ and $L=16$, respectively.
These results are compared with their exact counterparts.
We observe that the von Neumann entropy $S_t$ grows monotonically in time $t$ until it reaches the maximum value at $Jt=3.68$ (4.74) for $L=11$ (16). 
For time $t$ smaller than the maximum evolution time $t^\ast$ determined by $1-\mathcal{F}_t<10^{-2}$, the discrepancy between the von Neumann entropy $S_t$ and the exact value is as small as $10^{-2}$ for all gate sizes.

To further examine the entanglement structure, we also analyze the subsystem size dependence of the von Neumann entropy $S_t$ at time $t$, particularly when the exact value, calculated by partitioning the system in half, reaches its maximum.
Vertical dotted lines in Figs.~\ref{fig:compare-multi-qubit-l11-l16}(c) and \ref{fig:compare-multi-qubit-l11-l16_2}(c) indicate this maximum . 
As shown in Figs.~\ref{fig:compare-multi-qubit-l11-l16}(d) and \ref{fig:compare-multi-qubit-l11-l16_2}(d), the von Neumann entropy calculated for the exact time-evolved quantum state $|\Psi(t)\rangle$ forms a mountain-like shape as the subsystem size varies.
Introducing the approximation by gradually reducing the gate size $l$, the von Neumann entropy at the mountain's peak decreases while it still maintains its exact value to some extent for smaller subsystem sizes.
Therefore, to accurately describe the long-time dynamics using the approximate state $|\Phi(t)\rangle$ with the single-layer sequential-type quantum circuit composed of multi-qubit gates, the gate size $l$ of the multi-qubit gate should be increased.
This ensures a sufficiently large bond dimension, even when the system is partitioned into halves.

\section{\label{sec:diamond} Diamond-type decomposition}

In the previous section, we demonstrate that a quantum circuit consisting of multi-qubit gates well approximates the global quench dynamics.
However, to describe long-time dynamics, the gate size $l$ of the multi-qubit gate should be at least as large as half of the system size, i.e., $l\geq\lfloor L/2 \rfloor +1$. 
Therefore, this approach is not scalable because the $l$-qubit gate has $O(2^{2l})$ internal parameters, and thus the number of parameters increases exponentially with the system size $L$.
More exactly, since the $l$-qubit unitary gate with $p$ legs ($p<l$) connecting to the state $\ket{0}^{\otimes p}$ is described by the unitary matrix with $p$ qubit indices being fixed, i.e., 
\begin{align}
\hat{U}=&\sum_{\{\sigma_{j}\}}\sum_{\{\sigma'_{j}\}}
U^{\sigma_{1}\cdots\sigma_{l-p}\sigma_{l-p+1}\cdots\sigma_{l}}_{\sigma'_{1}\cdots\sigma'_{l-p}0\cdots 0}
\ket{\sigma_{1}}\bra{\sigma'_{1}}\otimes\cdots
\nonumber\\
&\otimes\ket{\sigma_{l-p}}\bra{\sigma'_{l-p}}
\otimes\ket{\sigma_{l-p+1}}\bra{0}\cdots\ket{\sigma_{l}}\bra{0},
\label{eq:unitary}
\end{align}
where $|\sigma_j\rangle$ with $\sigma_j=0,1$ is the eigenstate of Pauli operator $\hat{\sigma}_j^z$ at site $j$, i.e., 
$\hat{\sigma}_j^z|0\rangle=|0\rangle$ and $\hat{\sigma}_j^z|1\rangle=-|1\rangle$. 
The matrix $U^{\sigma_{1}\cdots\sigma_{l-p}\sigma_{l-p+1}\cdots\sigma_{l}}_{\sigma'_{1}\cdots\sigma'_{l-p}0\cdots 0}$ is composed of $2^{2l-p+1}$ independent real numbers with $2^{2l-2p}$ constraints due to its unitarity.
Hence, this unitary matrix is parametrized with $n_p=2^{2l-2p}(2^{p+1}-1)$ independent real numbers~\cite{lin_Real_2021}, which increases exponentially with the gate size $l$ even if $p=l-1$.
Thus, the multi-qubit gate-based quantum circuit with the large gate size $l$ is impractical when it is applied to a large system on real quantum computers.

\begin{figure}[htbp]
\includegraphics[width=0.9\columnwidth]{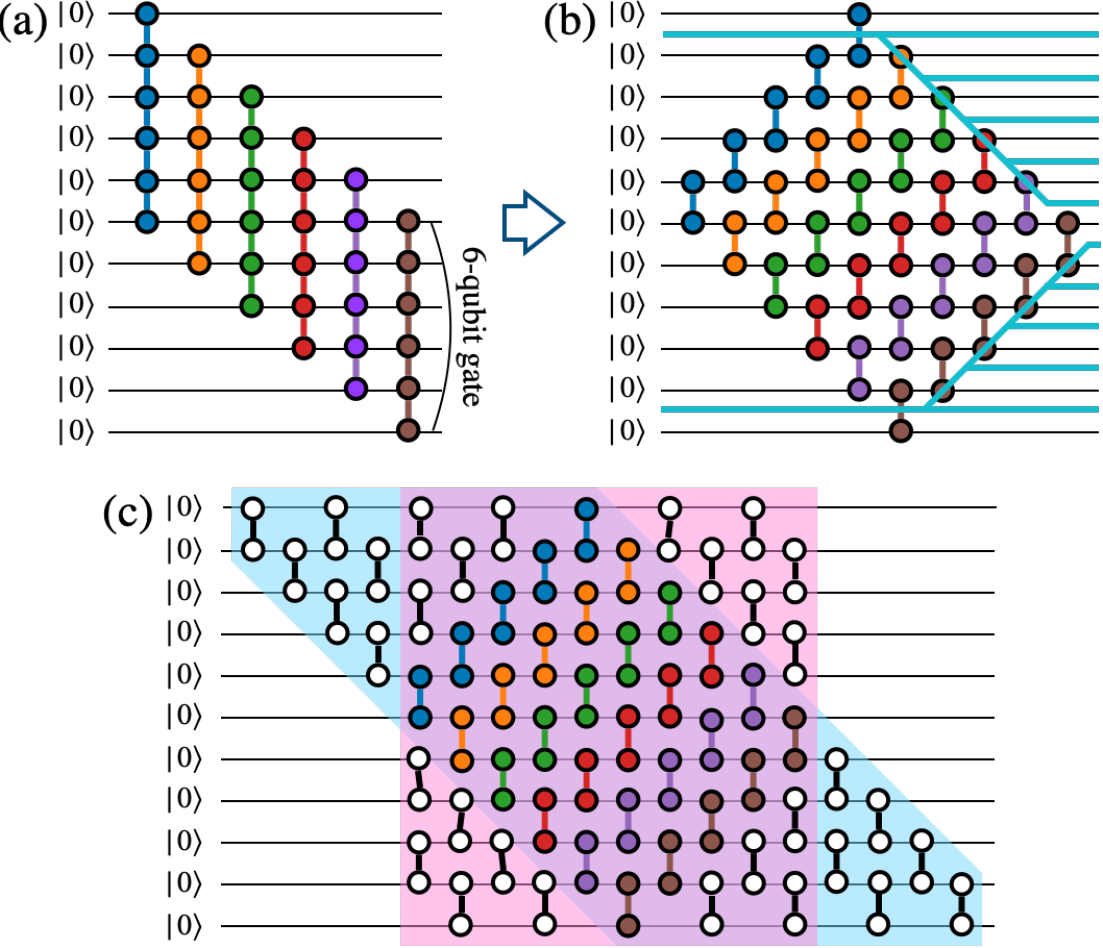}
\caption{
Decomposition of (a) a single-layer sequential-type quantum circuit composed of multi-qubit gates into (b) a diamond-shaped quantum circuit with two-qubit gates. 
Each multi-qubit gate indicated with a distinct color in (a) is approximately decomposed with a product of two-qubit gates denoted by the same color in (b). Notice that each set of two-qubit gates indicated with different colors in (b) is aligned sequentially. 
In this example, we consider that the system size is $L=11$ and the gate size of the multi-qubit gate is $l=6$, for which the single-layer sequential-type quantum circuit in (a) can describe exactly the global quench dynamics.  
The cyan solid lines in (b) denote examples of the shortest paths that separate the system into two subsystems, giving rise to the minimal bond dimension for each path in Eq.~(\ref{eq:Psi_w}), with the associated entanglement entropy being the upper limit (for more details, see Appendix~\ref{app:entanglement}).
(c) The diamond-shaped quantum circuit in (b) is included as a special case in both a brickwall-type quantum circuit of $5$ layers (indicated by the red-shaded region) and a five-layer sequential-type quantum circuit composed of two-qubit gates (indicated by the blue-shaded region).
}
\label{fig:diamond-decomposition}
\end{figure}

Because of the reason described above, we now consider the decomposition of a single-layer sequential-type quantum circuit composed of multi-qubit gates into the product of a finite number of two-qubit gates to reduce the internal parameters in a quantum circuit.
For this purpose, here we introduce a diamond-shaped quantum-circuit ansatz, which is the simplest approximation decomposing each multi-qubit gate in the sequential-type quantum circuit shown in Fig.~\ref{fig:diamond-decomposition}(a) into a sequential-type quantum circuit composed of two-qubit gates, as shown in Fig.~\ref{fig:diamond-decomposition}(b). 
It should be noted here that when the system is partitioned, the diamond-shaped quantum-circuit ansatz follows the volume law of the entanglement entropy since the shortest path dividing the system into two subsystems corresponds to the upper limit of the entanglement entropy (see Appendix~\ref{app:entanglement}).

First, we discuss the degrees of freedom included in several types of quantum circuits.
The number of two-qubit gates, $n_{g}$, in the diamond-shaped quantum circuit [see Fig.~\ref{fig:diamond-decomposition}(b)] is 
\begin{align}
n_{g}
=\begin{cases}
\frac{L^2-1}{4}&(L:\,\mathrm{odd})\\
\frac{L^2}{4}&(L:\,\mathrm{even})
\end{cases}, 
\end{align}
while the number of gates in the $m$-layer sequential-type and brickwall-type quantum circuits composed of two-qubit gates [see Fig.~\ref{fig:diamond-decomposition}(c)] are both given by $n_g=m(L-1)$. 
Hence, for example, when $L$ is odd, the number of two-qubit gates in the diamond-shaped quantum circuit for $L=4n+3$ ($n\in\mathbb{N}$) 
is exactly the same as those in the sequential-type and brickwall-type quantum circuits with $(n+1)$ layers for the same system size.
Moreover, as shown in Fig.~\ref{fig:diamond-decomposition}(c), the sequential-type and brickwall-type quantum circuits with 
$m=\lfloor L/2\rfloor$ layers contain the diamond-shaped quantum circuit.

Using this gate-counting rule, we can estimate the net number $n_p$ of independent real parameters in the diamond-shaped quantum circuit as follows.
\begin{align}
n_p=-1-4L+16\left\lfloor\frac{L}{2}\right\rfloor
\left(L-\left\lfloor\frac{L}{2}\right\rfloor\right)~.
\end{align}
In comparison, the number of parameters included in the single-layer sequential-type quantum circuit composed of multi-qubit gates with the gate size $l$ is given by:
\begin{align}
n_p=2^{l+1}-1+3(L-l)2^{2l-2}, 
\end{align}
where $l = \lfloor L/2 \rfloor + 1$ corresponds to the quantum circuit with the perfect expressivity. 
Therefore, the compression ratio of the internal degrees of freedom between these two quantum circuits for, e.g., $L=11$ is $15487/435 \approx 35.6$, and generally, it increases exponentially with increasing $L$. 
Thus, we can regard the diamond-shaped quantum circuit as a sparse approximation of the quantum circuit with multi-qubit gates.

\begin{figure}[htbp]
\includegraphics[width=1.0\columnwidth]{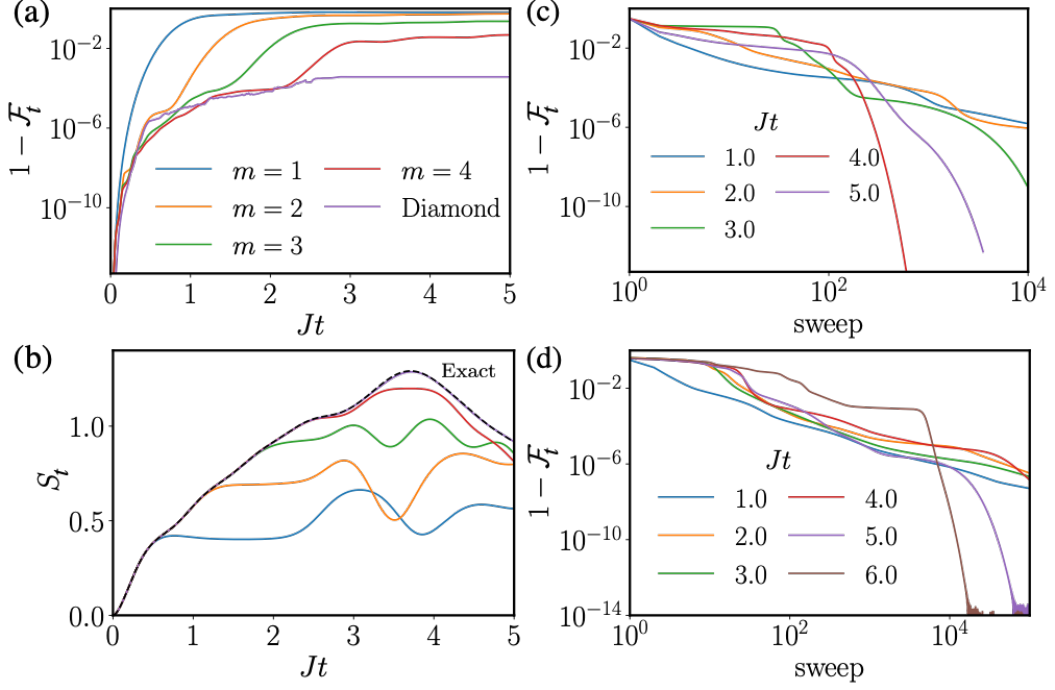}
\caption{
Time evolution of (a) the infidelity $1-\mathcal{F}_t$ and (b) the von Neumann entropy $S_t$ for the $m$-layer sequential-type quantum 
circuits consisting of two qubits and the diamond-shaped quantum circuit, approximating the time-evolved state $|\Psi(t)\rangle$ 
for the $L=11$ quantum Ising chain with the model parameters $g=1.4$ and $h=0$. 
The hyperparameters are set to be $w_\mathrm{max}=10^{4}$, $w_\mathrm{min}=10^{3}$, $\epsilon_{a}=10^{-14}$, 
and $\epsilon_{r}=10^{-4}$. 
For comparison, the exact result for $S_t$ is also shown by a dotted line in (b). 
(c) Sweep iteration dependence of the infidelity $1-\mathcal{F}_t$ evaluated for the optimized quantum circuit $\hat{C}_t$ 
with the diamond-shaped structure by directly referring to the exact time-evolved state 
$|\Psi(t)\rangle$ at $Jt=1, 2, \dots,5$ for the same quantum Ising model with the same hyperparameters as in (a) and (b).
(d) Same as (c) but for $L=16$, $w_\mathrm{max}=10^{5}$, and $w_\mathrm{min}=10^{4}$. 
}
\label{fig:evaluation-diamond}
\end{figure}

For benchmark calculations, Figs.~\ref{fig:evaluation-diamond}(a) and \ref{fig:evaluation-diamond}(b) illustrate respectively the time evolution of the infidelity $1-\mathcal{F}_t$ and the von Neumann entropy $S_t$ for the diamond-shaped quantum circuit and the $m$-layer sequential-type quantum circuits composed of two-qubit gates approximating global quench dynamics of the quantum Ising model with the longitudinal field $h=0$.
These results clearly show that the diamond-shaped quantum circuit is better than the sequential-type quantum circuit with $m\leq 4$ layers in terms of the expressivity of the exact time-evolved state $|\Psi(t)\rangle$ because the infidelity is well suppressed and the von Neumann entropy is in good agreement with the exact one, at least, in the time scale studied here. 
Since the number $n_g$ of two-qubit gates in the diamond-shaped quantum circuit is exactly the same as that in the three-layer sequential-type quantum circuit for the $L=11$ system, the diamond-shaped quantum circuit also performs better in compressibility than the sequential-type quantum circuits. 
For instance, for $Jt\geq 3$, the infidelity for the diamond-shaped quantum circuit is about $500$ times smaller than that for the three-layer sequential-type quantum circuit. 
Additionally, the infidelity for the diamond-shaped quantum circuit becomes almost constant for $Jt \geq 3$, implying that there is no error accumulation occurring in this time region, which is consistent with the results of the von Neumann entropy in Fig.~\ref{fig:evaluation-diamond}(b).

We also examine the accuracy of the diamond-shaped quantum circuit by optimizing the quantum circuit $\hat{C}_t$ to maximize the overlap $\mathcal{F}_{t}=\left|\langle\Psi(t)|\hat{C}_{t}|\Phi\rangle\right|^{2}$, directly referring to the exact time-evolved state $|\Psi(t)\rangle$.
As shown in Fig.~\ref{fig:evaluation-diamond}(c), we find that in the region of $Jt \geq 3$, the infidelity rapidly decreases after an appropriate number of sweep iterations for the optimization, and thus the quantum-circuit state $\ket{\Phi(t)}$ with the diamond-shaped structure can represent very accurately, i.e., numerically exactly, the exact time-evolved state $\ket{\Psi(t)}$.
This is consistent with the observation described above that the error does not accumulate in time, and this behavior is robust for larger systems up to $L=16$, as shown in Fig.~\ref{fig:evaluation-diamond}(d). 
It is also worth noting that a similar analysis for the three-layer sequential-type quantum circuit composed of two-qubit gates 
and the single-layer sequential-type quantum circuit composed of multi-qubit gates with the gate size $l=5$ does not 
find any optimized quantum circuits that reach the numerically exact solution 
%as in the case of the diamond-shaped quantum circuit 
(see Appendix \ref{app:optimize_each_timestep}).

\begin{figure}[htbp]
\includegraphics[width=1.0\columnwidth]{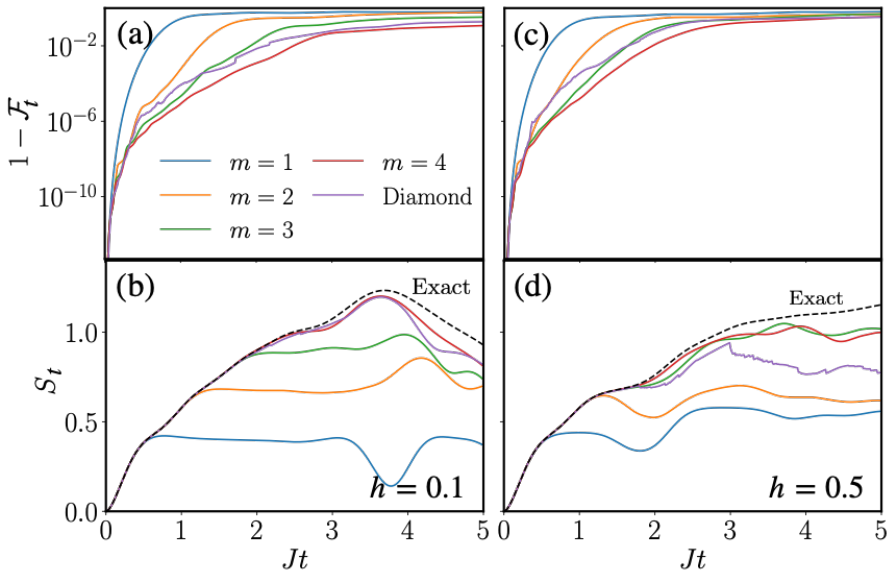}
\caption{
Time evolution of (a,c) the infidelity $1-\mathcal{F}_t$ and (b,d) the von Neumann entropy $S_t$ for the $m$-layer sequential-type 
quantum circuits consisting of two qubits and the diamond-shaped quantum circuit in the case of a finite longitudinal magnetic 
field $h=0.1$ (a,b) and 0.5 (c,d). Other model parameters and the hyperparameters for the circuit optimization are the same as those in 
Figs.~\ref{fig:evaluation-diamond}(a) and \ref{fig:evaluation-diamond}(b).
}
\label{fig:evaluation-diamond_h}
\end{figure}

However, as illustrated in Fig.~\ref{fig:evaluation-diamond_h}, the advantage of the diamond-shaped quantum circuit is gradually lost with increasing the longitudinal field $h$. 
A finite longitudinal field transforms the system from an integrable to a non-integrable one and enhances the complexity of quantum dynamics toward quantum information scrambling~\cite{karthik_Entanglement_2007,kim_Ballistic_2013}.
Although the diamond-shaped quantum circuit obeys the entropic volume law in its structure, there is an inherent difficulty in representing such quantum information scrambling with only very small internal degrees of freedom.
In contrast, the difficulties posed by the finite longitudinal field $h>0$ are absent when the single-layer sequential-type quantum circuit composed of multi-qubit gates is employed; the infidelity for the multi-qubit-gate-based quantum circuit monotonically decreases with increasing the size $l$ of multi-qubit gates, as already shown in Figs.~\ref{fig:compare-multi-qubit-l11-l16}(a) and \ref{fig:compare-multi-qubit-l11-l16_2}(a).

\begin{figure}[htbp]
\includegraphics[width=\columnwidth]{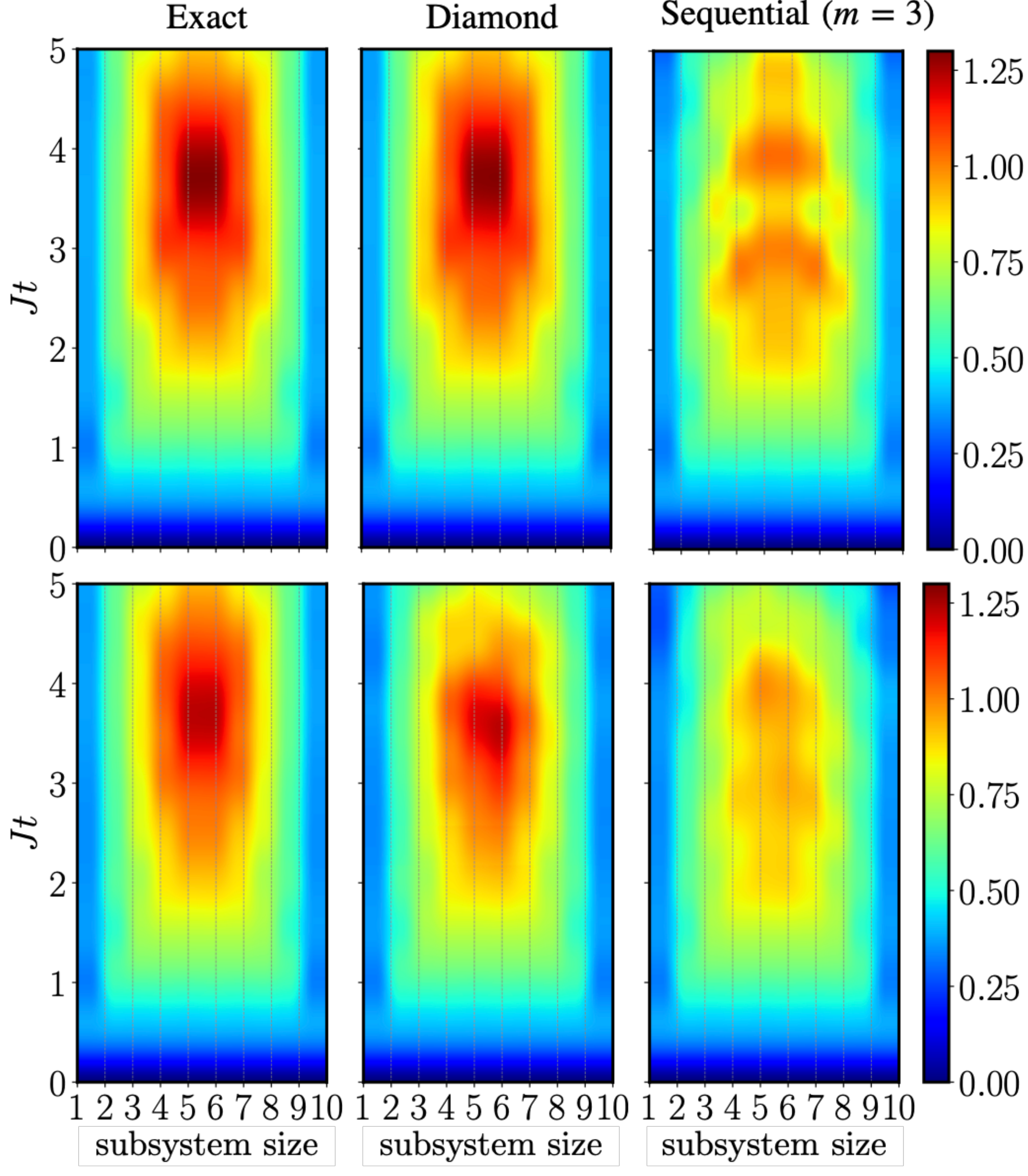}
\caption{The subsystem size dependence of the time evolution of the von Neumann entropy $S_t$ evaluated by the numerically exact calculation (left), the diamond-shaped quantum circuit (center), and the three-layer sequential-type quantum circuit consisting 
of two-qubit gates (right). 
The upper and lower panels are the results for the longitudinal field $h=0$ and $0.1$, respectively. 
Other model parameters and the hyperparameters for the circuit optimization 
are the same as those in Fig.~\ref{fig:evaluation-diamond_h}.}
\label{fig:evaluation-diamond-colormap}
\end{figure}

Finally, we examine in Fig.~\ref{fig:evaluation-diamond-colormap} the spacetime distribution of the von Neumann entropy 
$S_t$ evaluated for the diamond-shaped quantum circuit and the $m$-layer sequential-type quantum circuit with $m=3$,   
also compared with the exact one. 
Here, the quantum circuits are optimized by maximizing $I(t)$ in Eq.~(\ref{eq:i}). 
Initially, we observe the von Neumann entropy in the exact calculation forms a mountain-like shape in spacetime with a peak 
located near the central bond of the system at time $Jt=3.68$ for the longitudinal field $h=0$, while the von Neumann entropy 
at the edges almost does not increase in time.
This spacetime distribution of the von Neumann entropy is in good accordance with the results obtained for 
the diamond-shaped quantum circuit in which 
the depth of the quantum circuit increases linearly from the edge to the center of the system. 
On the other hand, the three-layer sequential-type quantum circuit containing the same number $n_g$ of two-qubit gates as the diamond-shaped quantum circuit needs longer circuit depth near the center of the system to capture the whole entanglement structure. 
This advantage of the diamond-shaped quantum circuit in describing the spacetime distribution of the von Neumann entropy is 
explicitly confirmed for the longitudinal field up to $h=0.1$, as shown in the lower panels of Fig.~\ref{fig:evaluation-diamond-colormap}, 
which is also consistent with the results in Fig.~\ref{fig:evaluation-diamond_h}.

\section{\label{sec:conclusion} Conclusion}

We have investigated the quantum-circuit ansatz that can efficiently describe the real-time dynamics of quantum states in the 1D quantum Ising model with the longitudinal and transverse fields after a global quantum quench.
First, we examined the expressiveness of the single-layer sequential-type quantum circuit composed of multi-qubit gates with the gate size $l$. 
Our numerical calculations utilizing this multi-qubit-gate-based quantum circuit revealed that the evolution time of the quench dynamics for which the quantum-circuit ansatz can faithfully represent the time-evolved state within a certain accuracy increases almost linearly in the gate size $l$. 
Moreover, based on the analysis of the tensor-network representation of the quantum circuits, 
we showed that the single-layer sequential-type quantum circuit composed of multi-qubit gates with the gate size $l>L/2$ can represent exactly any quantum state, which was also confirmed numerically.

In order to reduce the internal degrees of freedom in the multi-qubit-gate-based quantum circuit, we subsequently explored two distinct types of quantum circuits, i.e., the diamond-shaped quantum circuit and the multi-layer sequential-type quantum circuit constructed using two-qubit gates. 
The diamond-shaped quantum circuit contains $O(L^2)$ real parameters as its internal degrees of freedom, following the volume law of entanglement, and is encompassed within the multi-layer sequential-type quantum circuit with $m=\lfloor L/2 \rfloor$ layers.
As a benchmark, we have examined the accuracy of these two types of quantum circuits in describing the long-time evolution of the quench dynamics in the transverse field Ising model without the longitudinal field.
Our numerical calculations found that the infidelity of the diamond-shaped quantum circuit is approximately 500 times smaller than that of the three-layer sequential-type quantum circuit, which contains the same number of two-qubit gates as the diamond-shaped quantum circuit. 
Intriguingly, we found that the diamond-shaped quantum circuit achieves smaller infidelity than the single-layer sequential-type quantum circuits composed of multi-qubit gates with the gate size $l\leq \lfloor L/2 \rfloor$, demonstrating an advantage in the sparsity of the diamond-shaped quantum circuit that satisfies the entropic volume law for time-evolved states in non-equilibrium quantum dynamics.

However, it should be noted that the diamond-shaped quantum circuit does not invariably succeed in describing time-evolved quantum states. 
In fact, for the quench dynamics of the same quantum Ising model but with a finite longitudinal field $h>0$, which amplifies the quantum information scrambling, no difference in the expressivity is found between the diamond-shaped quantum circuit and the multi-layer sequential-type quantum circuits, both of which do not successfully describe the long-time evolution of the quantum dynamics 
specially when $h>0.1$. 
On the other hand, no difficulty is observed in describing the long-time dynamics even with a finite longitudinal field $h>0$ when the multi-qubit-gate-based quantum circuit is employed.
Therefore, these studies highlight the importance of not only the entropic volume law but also the inherent degrees of freedom in the quantum circuit to describe long-time dynamics in general.

One of the promising avenues for future studies is the classification of different quantum dynamical systems based on their compatible quantum circuits for describing these systems, such as the diamond-shaped, sequential-type, and brickwall-type quantum circuits.
A noteworthy example is that the diamond-shaped quantum circuit is highly effective in describing time-evolved states of the quantum Ising model with the longitudinal field $h=0$, which can plausibly be attributed to the one-body nature of the system.
The number of parameters necessary to describe the time-evolution operator for a noninteracting one-body fermion model is considerably fewer than an interacting many-body fermion model, and the 1D transverse-field Ising model can be transformed into the one-body fermionic Hamiltonian using the Jordan-Wigner transformation~\cite{jordan_Uber_1928}. 
Moreover, it has been established that a quantum circuit with a depth of layers less than the system size can exactly describe the real-time dynamics of free fermions~\cite{shirakawa_Discretized_2021}. 
Therefore, it is highly interesting to extend these analyses to a perturbative region near free fermion systems in light of quantum dynamics.

A significant challenge is the optimization procedure to obtain an optimized quantum circuit with a given circuit structure 
for describing time-evolved states. 
Our numerical calculations have revealed that while the optimization of the single-layer sequential-type quantum circuit composed of $l$-qubit gates can typically be completed within at most $100$ sweep iteration updates, regardless of the gate size $l$, the number of sweep iteration updates required is typically $10-100$ times larger once the number of layers is increased, even for $l=2$. 
In particular, the numerical cost for a larger system becomes more expensive. 
Therefore, it is imperative to improve numerical algorithms that can optimize the multi-layer quantum circuits faster by, e.g., adjusting the initial conditions of the quantum circuit and the order of unitary gates to be optimized.

\begin{acknowledgments}
This work is supported by Grant-in-Aid for Scientific Research (A) (No.~JP21H04446), 
Grant-in-Aid for Scientific Research (B) (No.~JP21H03455 and No.~JP22H01171) 
and Grant-in-Aid for Scientific Research (C) (No.~JP22K03479) from MEXT, Japan, 
and Grant-in-Aid for Transformative Research Areas "The Natural Laws of Extreme
Universe---A New Paradigm for Spacetime and Matter from Quantum Information"
(No.~JP21H05182 and No.~JP21H05191) from JSPS, Japan.
It is also supported by JST PRESTO (No.~JPMJPR1911), MEXT Q-LEAP (No.~JPMXS0120319794), 
and JST COI-NEXT (No.~JPMJPF2014 and No.~JPMJPF2221). 
It is also partially supported by Program for Promoting Research on the Supercomputer 
Fugaku (No.~JPMXP1020230411) form MEXT, Japan, and by the COE research grant 
in computational science from Hyogo Prefecture and Kobe City through Foundation for Computational Science.
We are grateful for allocating computational resources of the HOKUSAI BigWaterfall supercomputing system at RIKEN and SQUID at the Cybermedia Center, Osaka University.
\end{acknowledgments}

\appendix

\section{\label{app:method} Local Optimization Algorithm}

\subsection{Quantum circuit encoding algorithm}

We first briefly summarize the quantum circuit encoding (QCE) algorithm originally proposed in Ref.~\cite{shirakawa_Automatic_2021}.
Let us assume that $\ket{\Psi}$ is a target quantum state and construct a variational state $\ket{\Phi}=\hat{C}\ket{0}^{\otimes L}$ which approximates the target state $\ket{\Psi}$.
Here, $\ket{0}^{\otimes L}$ is the direct product of local states $|0\rangle$ for the $L$-site system and $\hat{C}$ is the quantum circuit defined as
\begin{align}
\hat{C}=\hat{U}_{N}\cdots \hat{U}_{1}\equiv\prod^{1}_{i=N}\hat{U}_{i}, 
\end{align}
where each unitary operator $\hat{U}_{i}$ acts on $l_i\,(<L)$ qubits expanded by the orthonormal set $\{ \ket{{\sigma}_{b_{1}} \cdots {\sigma}_{b_{l_i}}} \}$ with $1 \leq b_{1} < \cdots < b_{l_i} \leq L$.
The index pool $B_{i}=\{b_{1},b_{2},\cdots,b_{l_i}\} \in \mathbb{B}$ for specifying a subsystem is a hyperparameter of the QCE algorithm~\cite{shirakawa_Automatic_2021}.
However, in this study, the form of the quantum circuit, i.e., the gate sets as well as the loci that the gates act on, is fixed.

The QCE algorithm seeks an optimal quantum circuit $\hat{C}$ to maximize the overlap between the two states $\ket{\Psi}$ and $\ket{\Phi}$, i.e.,  $F=\mathrm{Re}\braket{\Phi|\Psi}=\mathrm{Re}\bra{0}\hat{C}^{\dagger}\ket{\Psi}$.
Now, let us focus on the update of $i$-th unitary operator $\hat{U}_{i}$ in the quantum circuit $\hat{C}$.
Introducing $\bra{\Phi_{i}}=\bra{0}\prod^{i-1}_{j=1}\hat{U}_{j}^{\dagger}$ and $\ket{\Psi_{i}}=\prod^{L}_{j=i+1}\hat{U}_{j}^{\dagger}\ket{\Psi}$, we can rewrite $F$ as follows:
\begin{align}
F=\mathrm{Re} \mathrm{Tr}\left[\ket{\Psi_{i}}\bra{\Phi_{i}}U_{i}^{\dagger}\right]
=\mathrm{Re} \mathrm{Tr}_{B_{i}} \left[\hat{F}_{i}\hat{U}_{i}^{\dagger}\right],
\label{eq:QCE-fidelity}
\end{align}
where $\hat{F}_{i}=\mathrm{Tr}_{\bar{B}_{i}}\ket{\Psi_{i}}\bra{\Phi_{i}}$ and $\bar{B}_{i}$ denotes the complement 
of the subsystem $B_{i}$.
Then, applying the singular value decomposition (SVD) to $\hat{F}_{i}$ as $\hat{F}_{i} = \hat{X}_{i} \hat{D}_{i} \hat{Y}_{i}^{\dagger}$ with unitary operators $\{ \hat{X}_{i}, \hat{Y}_{i} \}$ and the diagonal non-negative operator $\hat{D}_{i}$, we can further rewrite $F$ as
\begin{align}
F=\mathrm{Re}\mathrm{Tr}_{B_{i}} \left[\hat{D}_{i} \hat{Y}_{i}^{\dagger} \hat{U}_{i}^{\dagger} \hat{X}_{i} \right]
\le \mathrm{Tr}_{B_{i}} \hat{D}_{i}, 
\end{align}
because the norm of diagonal elements for the unitary $\hat{Y}_{i}^{\dagger} \hat{U}_{i}^{\dagger} \hat{X}_{i}$ 
is less than or equal to one.
Therefore, updating $\hat{U}_{i}^{\dagger}=\hat{Y}_{i} \hat{X}_{i}^{\dagger}$, we can locally maximize $F$. 
Note that in Ref.~\cite{shirakawa_Automatic_2021}, instead of maximizing $F$, the quantity $|\braket{\Phi|\Psi}|$ is maximized. However, the resulting unitary operator $\hat{U}_i^\dag$ maximizing this quantity is exactly the same as that obtained here. 

Within the QCE algorithm, the same update is repeated with incrementing $i$ form $i=1$. 
After updating $\hat{U}_{i}$ up to $i=N$, we turn around the update order and decrement $i$ down to $i=1$. 
This sequence of updates is referred to as one sweep update. The QCE algorithm performs these sweep updates several times 
until $F$ converges.

\subsection{Time-evolution algorithm}

In quantum mechanics, the dynamics for the time-dependent Hamiltonian $\hat{H}(t)$ can be described by the time-dependent 
Schr\"odinger equation  
\begin{align}
i\frac{\partial}{\partial t}\ket{\Psi(t)}=\hat{H}(t)\ket{\Psi(t)}
\end{align}
with the time-evolved state $\ket{\Psi(t)}$ and the reduced Planck constant $\hbar=1$ .
This equation can be formally resolved as
\begin{align}
&\ket{\Psi(t)}=\hat{U}(t,t_{0})\ket{\Psi(t_{0})}
\end{align}
with
\begin{align}
&\hat{U}(t,t')=T_{t}\left[ e^{-{i}\int^{t}_{t'}dt''~\hat{H}(t'')} \right],
\end{align}
where $T_{t} \left[ \cdot \right]$ denotes the time-ordering product.
When we split the time into $N$ pieces, i.e., $t_{n}=n\Delta t+t_{0}$, with 
$\Delta t=\frac{t-t_{0}}{N}$ and $n=0,1,\dots,N$, we can obtain the recurrence formula:
\begin{align}
\hat{U}(t_{n+1},t_{n})\ket{\Psi(t_{n})}=\ket{\Psi(t_{n+1})}.
\end{align}

If a parametrized quantum circuit $\hat{C}_{n} = \hat{C}({\bm \theta}_n)$ has sufficiently large degrees of freedom ${\bm \theta}_n$ and can represent the dynamical state at time $t_n$, namely, $\ket{\Psi(t_{n})}=\hat{C}_{n}\ket{0}$, this recurrence formula can be rewritten 
for the time-evolved state $|\Psi(t_{n+1})\rangle $ at time $t_{n+1}$ as
\begin{align}
e^{-i\Delta t \hat{H}(t_{n})}\hat{C}_{n}\ket{0}=\hat{C}_{n+1}\ket{0}
\end{align}
in the small $\Delta t$ limit.
In this case, we can formulate the time-dependent variational principle by minimizing the following cost function:
\begin{align}
\tilde{F}^{(n+1)}=&\left\|e^{-i\Delta t H}\hat{C}_{n}\ket{0}-\hat{C}_{n+1}\ket{0}\right\|^2
\nonumber\\
=&2-2\mathrm{Re}\bra{0}\hat{C}_{n+1}^{\dagger}e^{-i\Delta t H}\hat{C}_{n}\ket{0}.
\label{eq:Fcost}
\end{align}
Therefore, we can construct an optimal quantum circuit $\hat{C}_{n+1}$ for $|\Psi(t_{n+1})\rangle $ by solving the following equation: 
\begin{align}
\hat{C}_{n+1}=\argmax_{\hat{C}}\mathrm{Re}\bra{0}\hat{C}^{\dagger}e^{-i\Delta t H}\hat{C}_{n}\ket{0}, 
\end{align}
which is equivalent to maximizing $I(t+\Delta t)$ in Eq.~(\ref{eq:i}). 
Continuing this procedure iteratively, we can investigate the real-time dynamics of the parametrized quantum circuit.

\section{\label{app:exact-circuit} Exact Quantum Circuit Representation}

This appendix explains that an arbitrary quantum state can be explicitly written as a single-layer sequential-type quantum circuit 
composed of multi-qubit gates with the qubit size $l=\lfloor L/2 \rfloor+1$ (see Fig.~\ref{fig:circuit-multi-qubit}).
Let us first introduce the normalized quantum state
\begin{align}
\ket{\Psi}=\sum_{\{\sigma_{j}\}}c_{\sigma_{1}\cdots\sigma_{L}}
\ket{\sigma_{1}\cdots\sigma_{L}},
\label{eq:psi}
\end{align}
where $L$ is the system size and $\sigma_j \in \{0, 1\}$ with $1 \leq j \leq L$ represents the local state at the $j$th qubit.

Now, we consider $\{\sigma_1 \cdots \sigma_{L-l}\}$ and $\{ \sigma_{L-l+1} \cdots \sigma_{L} \}$ to be indices specifying the row and 
column components of $c_{\sigma_{1}\cdots\sigma_{L}}$ in Eq.~(\ref{eq:psi}), and apply the SVD to this matrix as 
\begin{align}
c_{\sigma_{1}\cdots\sigma_{L}}=&
\sum_{\lambda_{L-l}=0}^{2^{L-l}-1}
Y^{[L-l],\sigma_{1}\cdots\sigma_{L-l}}_{\lambda_{L-l}} D^{[L-l]}_{\lambda_{L-l}}
B^{\sigma_{L-l+1}\cdots\sigma_{L}}_{\lambda_{L-l}} \nonumber \\
=&
\sum_{\lambda_{L-l}=0}^{2^{L-l}-1}
X^{[L-l],\sigma_{1}\cdots\sigma_{L-l}}_{\lambda_{L-l}} B^{\sigma_{L-l+1}\cdots\sigma_{L}}_{\lambda_{L-l}} 
\label{eq:x}
\end{align}
with $X^{[L-l],\sigma_{1}\cdots\sigma_{L-l}}_{\lambda_{L-l}} = Y^{[L-l],\sigma_{1}\cdots\sigma_{L-l}}_{\lambda_{L-l}} D^{[L-l]}_{\lambda_{L-l}}$. Notice that here we use $L-l<l$ for $l=\lfloor L/2 \rfloor+1$ and thus the upper bound of the sum above for 
$\lambda_{L-l}$ is $2^{L-l}-1$.  
In addition, the left-singular vectors $\{ Y^{[L-l],\sigma_{1}\cdots\sigma_{L-l}}_{\lambda_{L-l}} \}$, 
the right-singular vectors $\{ B^{\sigma_{L-l+1}\cdots\sigma_{L}}_{\lambda_{L-l}} \}$, and 
the singular values $\{ D^{[L-l]}_{\lambda_{L-l}} \}$ satisfy, respectively, the following relations: 
\begin{align}
&
\sum_{\sigma_1,\, \cdots,\, \sigma_{L-l}} \left( Y^{[L-l],\sigma_{1}\cdots\sigma_{L-l}}_{\lambda_{L-l}} \right)^* Y^{[L-l],\sigma_{1}\cdots\sigma_{L-l}}_{\lambda'_{L-l}} = 
\delta_{\lambda^{~}_{L-l},\lambda'_{L-l}} \nonumber \,, \\&
\sum_{\sigma_{L-l+1},\, \cdots,\, \sigma_{L}} \left( B^{\sigma_{L-l+1} \cdots \sigma_{L}}_{\lambda_{L-l}} \right)^* B^{\sigma_{L-l+1} \cdots \sigma_{L}}_{\lambda'_{L-l}} = 
\delta_{\lambda^{~}_{L-l},\lambda'_{L-l}} \nonumber \,, \\&
\sum_{\lambda_{L-l}} \left( D^{[L-l]}_{\lambda_{L-l}} \right)^2 = 1 
\end{align}
with $\delta_{\lambda^{~}_{L-l},\lambda'_{L-l}}$ being the Kronecker delta.

Next, we consider $\{\sigma_1 \cdots \sigma_{L-l-1}\}$ and $\{ \sigma_{L-l} \lambda_{L-l} \}$ to be indices specifying the row and 
column components of $X^{[L-l],\sigma_{1} \cdots \sigma_{L-l}}_{\lambda_{L-l}}$ in Eq.~(\ref{eq:x}), and apply 
the SVD to this matrix to obtain 
\begin{align}
X^{[L-l],\sigma_{1} \cdots \sigma_{L-l}}_{\lambda_{L-l}} = & \!\!\!\!\!
\sum_{\lambda_{L-l-1}=0}^{2^{L-l-1}-1} \!\!\!\!\!
X^{[L-l-1],\sigma_{1}\cdots\sigma_{L-l-1}}_{\lambda_{L-l-1}} A^{[L-l-1],\sigma_{L-l}}_{\lambda_{L-l-1},\lambda_{L-l}} ,~
\end{align}
where the right-singular vectors $\{ A^{[L-l-1],\sigma_{L-l}}_{\lambda_{L-l-1},\lambda_{L-l}} \}$ satisfies
\begin{align}
\sum_{\sigma_{L-l}, \lambda_{L-l}} \!\!\!
\left( A^{[L-l-1],\sigma_{L-l}}_{\lambda_{L-l-1},\lambda_{L-l}} \right)^* A^{[L-l-1],\sigma_{L-l}}_{\lambda'_{L-l-1},\lambda_{L-l}} = 
\delta_{\lambda^{~}_{L-l-1},\lambda'_{L-l-1}} \nonumber ~.
\end{align}

Then, recursively executing this procedure, we can finally rewrite $c_{\sigma_{1}\cdots\sigma_{L}}$ as the following MPS form:
\begin{align}
c_{\sigma_{1}\cdots\sigma_{L}}=&
\sum_{\lambda_1=0}^1\cdots\sum_{\lambda_{L-1}=0}^{2^{L-l}-1}A^{[1],\sigma_{1}}_{0,\lambda_{1}}
\cdots A^{[L-l],\sigma_{L-l}}_{\lambda_{L-l-1},\lambda_{L-l}}
B^{\sigma_{L-l+1}\cdots\sigma_{L}}_{\lambda_{L-l}}.
\label{eq:mps_c}
\end{align}

\begin{figure}[htbp]
\includegraphics[width=\columnwidth]{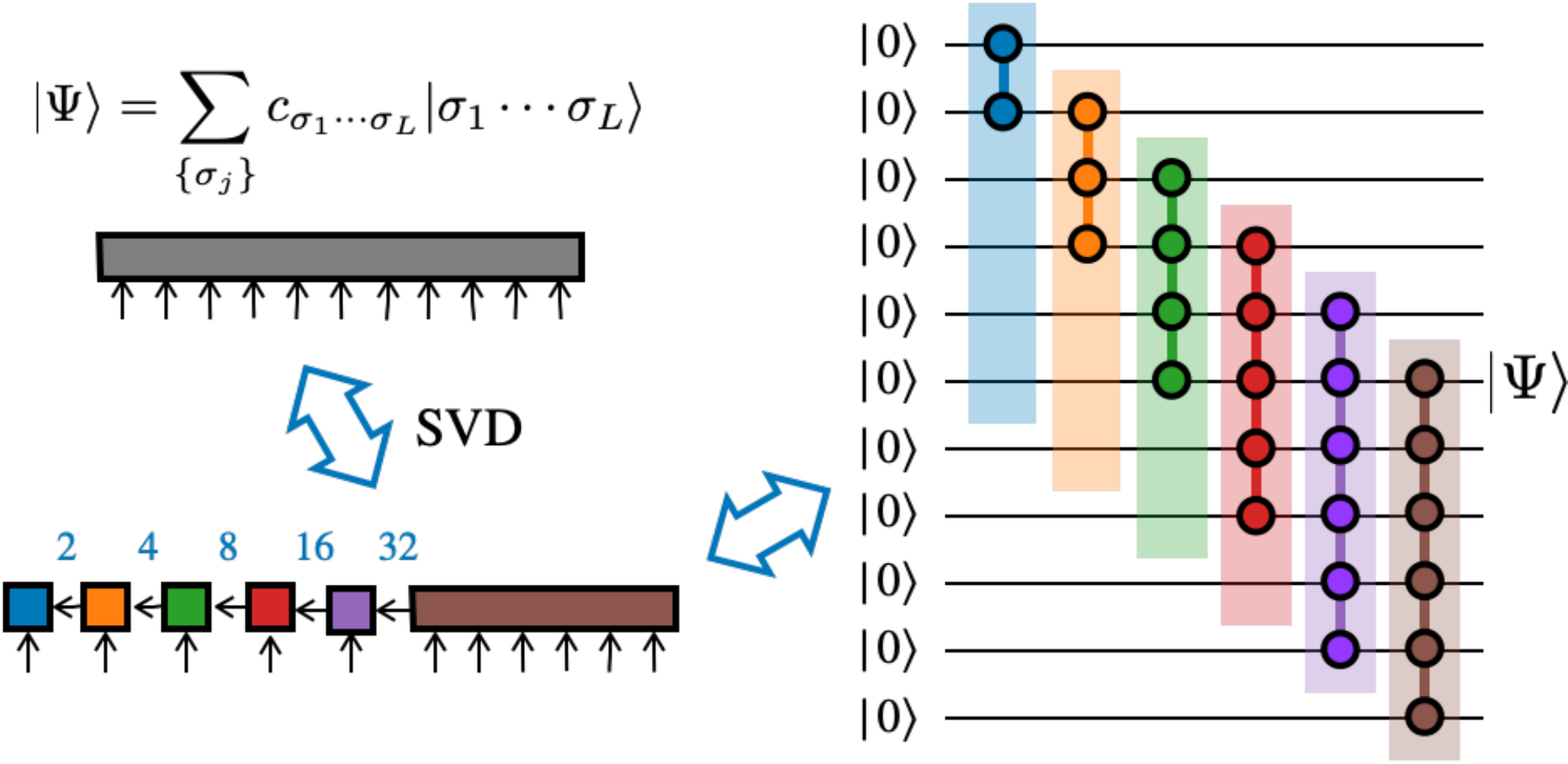}
\caption{
The schematic figure of the transformation between the exact MPS representation  
(the left bottom panel) 
and the exact quantum-circuit representation (the right panel) for a quantum state $|\Psi\rangle$ (the left top panel). 
The numbers in the left bottom panel denote the bond dimensions of the corresponding
links in the MPS. In the right panel, 
the differently colored dumbbell-like objects represent different multi-qubit gates describing the unitary operators 
$\hat{U}_i$ in Eq.~(\ref{eq.psi_u}), and 
the shaded regions indicate that the exact quantum-circuit representation
can be embedded into the single-layer sequential-type quantum circuit composed of multi-qubit gates with the gate size 
$l=\lfloor\frac{L}{2}\rfloor+1$. In this example, we consider the system size $L=11$.
}
\label{fig:exact-decomposition}
\end{figure}

Let us now introduce unitary operators representing a $(i+1)$-qubit gate 
for $1 \leq i \leq L-l$, 
\begin{align}
\hat{U}_{i}=&\sum_{\sigma_i,\, \cdots,\, \sigma_{2i}} \sum_{\sigma'_i,\, \cdots,\, \sigma'_{2i}}
[U_{i}]^{\sigma_{i}\cdots\sigma_{2i}}_{\sigma'_{i}\cdots\sigma'_{2i}}
\ket{\sigma_{i}\cdots\sigma_{2i}}\bra{\sigma'_{i}\cdots\sigma'_{2i}},
\end{align}
and a unitary operator representing a $l$-qubit gate, 
\begin{align}
\hat{U}_{L-l+1}=&\sum_{\sigma_{L-l+1},\, \cdots,\, \sigma_{L}} \sum_{\sigma'_{L-l+1},\, \cdots,\, \sigma'_{L}}
[U_{L-l+1}]^{\sigma_{L-l+1} \cdots \sigma_{L}}_{\sigma'_{L-l+1} \cdots \sigma'_{L}} \nonumber \\ &
\times \ket{\sigma_{L-l+1} \cdots \sigma_{L}}\bra{\sigma'_{L-l+1} \cdots \sigma'_{L}}.
\end{align}
Then, we can straightforwardly embed elements of tensors constituting the MPS in Eq.~(\ref{eq:mps_c})
into the unitary matrices by rewriting the bond index $\lambda_{i}$ as the binary number $\lambda_{i}=(x_{1}\cdots x_{i})$, i.e.,  
\begin{align}
&[U_{i}]^{\sigma_i x'_{1}\cdots x'_{i}}_{x_{1}\cdots x_{i-1}00}
= A^{[i],\sigma_i}_{(x_{1}\cdots x_{i-1}),(x'_{1}\cdots x'_{i})} 
\end{align}
for $1 \leq i \leq L-l$ and 
\begin{align}
&[U_{L-l+1}]^{\sigma_{L-l+1}\cdots\sigma_{L}}_{x_{1}\cdots x_{L-l}0\cdots 0}
= B^{\sigma_{L-l+1}\cdots\sigma_{L}}_{(x_{1}\cdots x_{L-l})}\,.
\end{align}
Using these unitary operators, the MPS representation of $\ket{\Psi}$ can be transformed into the following quantum circuit 
representation (also see Fig.~\ref{fig:exact-decomposition}):  
\begin{align}
\ket{\Psi}=\hat{U}_{L-l+1}\hat{U}_{L-l}\cdots\hat{U}_{1}\ket{0}^{\otimes L}. 
\label{eq.psi_u}
\end{align}
Therefore, we can easily confirm that the single-layer sequential-type quantum circuit composed of multi-qubit gates with the gate size $l$ contains this quantum circuit (see the right panel in Fig.~\ref{fig:exact-decomposition}).

\begin{figure}[htbp]
\includegraphics[width=1.0\columnwidth]{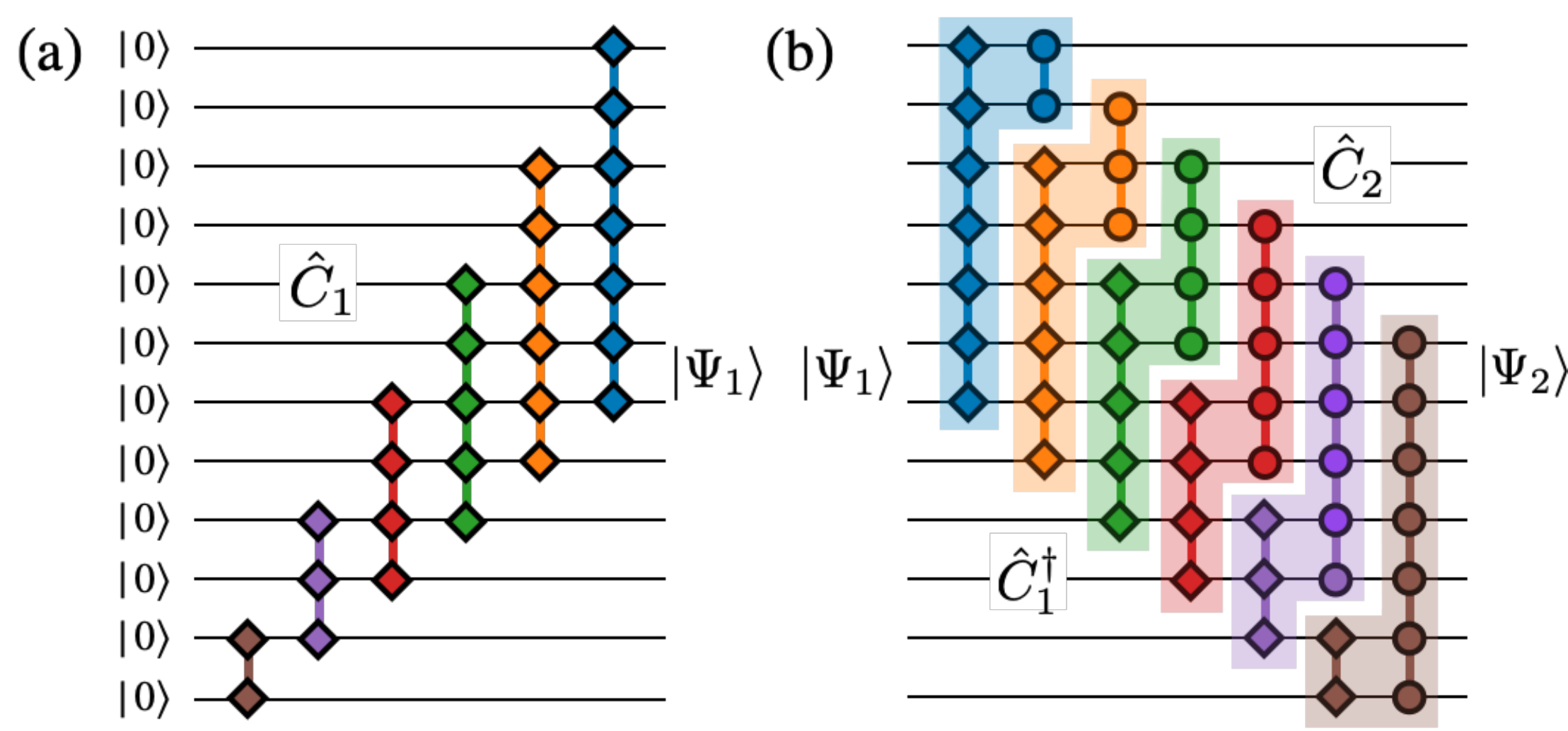}
\caption{
(a) Exact quantum circuit representation of a quantum state $|\Psi_1\rangle$, i.e., $|\Psi_1\rangle=\hat{C}_1|0\rangle$, which 
is essentially the same as that shown in Fig.~\ref{fig:exact-decomposition} but is composed of multi-qubit gates aligned 
differently from the bottom left to the top right. 
(b) Quantum circuit representation that transforms a quantum state $|\Psi_1\rangle$ to a quantum state $|\Psi_2\rangle$, i.e., 
$|\Psi_2\rangle=\hat{C}|\Psi_1\rangle$. The unitary transformation $\hat{C}$ is easily constructed as 
$\hat{C}=\hat{C}_2\hat{C}_1^\dag$, where $\hat{C}_1$ and $\hat{C}_2$ are obtained exactly as in (a) and the right panel 
in Fig.~\ref{fig:exact-decomposition}, respectively. 
Since the two multi-qubit gates indicated by the same color are combined into a single multi-qubit gate (shaded color) with the 
gate size $l=\lfloor\frac{L+1}{2}\rfloor+1$, this proves that the unitary transformation $\hat{C}$ representing the transition between 
arbitrarily two quantum states $|\Psi_1\rangle$ and $|\Psi_2\rangle$ can be described by a single-layer sequential-type quantum circuit 
composed of $l$-qubit gates. The example here considers the system size $L$ even (i.e., $L=12)$, but the same statement is also correct for $L$ odd. 
}
\label{fig:exact-decomposition-unitary}
\end{figure}

Finally, we prove that the unitary transformation $\hat{C}$ representing the transition between two normalized states $\ket{\Psi_{1}}$ 
and $\ket{\Psi_{2}}$, i.e., $|\Psi_{2}\rangle=\hat{C}|\Psi_{1}\rangle$, can be described by a single-layer sequential-type quantum circuit 
with $l(=\lfloor\frac{L+1}{2}\rfloor+1)$-qubit gates. 
For this purpose, we first note that these two states $\ket{\Psi_{1}}$ and $\ket{\Psi_{2}}$ can be described exactly as 
$\ket{\Psi_{1}}=\hat{C}_{1}|0\rangle^{\otimes L}$ and $\ket{\Psi_{2}}=\hat{C}_{2}|0\rangle^{\otimes L}$, where $\hat{C}_1$ and 
$\hat{C}_2$ are given respectively by two distinct sets of $L-l+1$ unitary operators as in Eq.~(\ref{eq.psi_u}) (also see the right panel in Fig.~\ref{fig:exact-decomposition}). 
However, here we assume that $\hat{C}_1$ is now composed of the unitary operators representing multi-qubit gates aligned as in Fig.~\ref{fig:exact-decomposition-unitary}(a), which is legitimately always possible simply because one can also divide the indices of $c_{\sigma_1\cdots\sigma_L}$ in Eq.~(\ref{eq:x}) into $\{\sigma_1\cdots\sigma_l\}$ and $\{\sigma_{l+1}\cdots\sigma_L\}$, and follow the similar analysis given in Eqs.~(\ref{eq:x})--(\ref{eq.psi_u}) in order to decompose $c_{\sigma_1\cdots\sigma_L}$ from the left, thus the left-singular vectors being now implemented in the unitary operators.
The unitary transformation $\hat{C}$, simply given by $\hat{C}=\hat{C}_2\hat{C}_1^\dag$, is thus schematically shown in Fig.~\ref{fig:exact-decomposition-unitary}(b). From this figure, it is obvious that $\hat{C}$ is described by a 
single-layer sequential-type quantum circuit composed of multi-qubit gates with the gate size $l=\lfloor\frac{L+1}{2}\rfloor+1$. 
We have also numerically checked that $\hat{C}$ can be always obtained exactly within the numerical accuracy 
for randomly generated states $|\Psi_1\rangle$ and $|\Psi_2\rangle$ by optimizing the single-layer sequential-type quantum circuit 
with $l$-qubit gates. 

\section{\label{app:entanglement} Upper Bound of Entanglement Entropy for Quantum-Circuit States}

\begin{figure*}[htbp]
\includegraphics[width=2.\columnwidth]{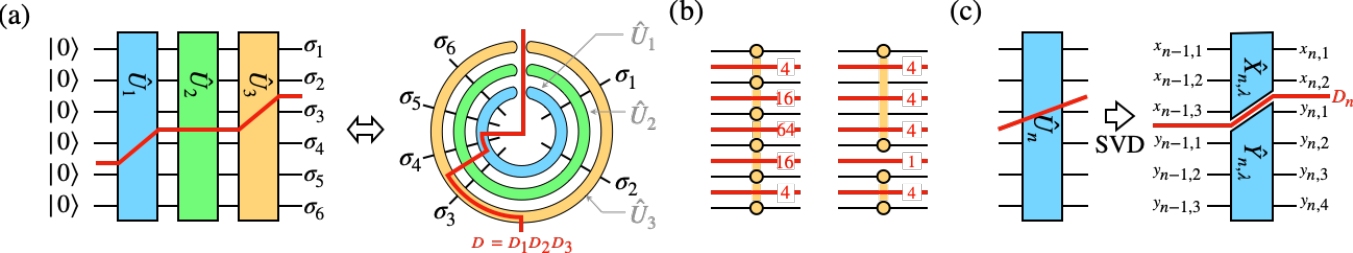}
\caption{
(a) Left panel: the quantum-circuit state $|\Psi\rangle$ composed of 3-stacked unitary gates for $L=6$. The red line indicates 
an example of paths for 1D bipartition $\{\sigma_{1}\sigma_{2}|\sigma_{3}\sigma_{4}\sigma_{5}\sigma_{6}\}$. 
Right panel: deformation of the quantum-circuit state in the left panel with $\{\sigma_{1}\sigma_{2}\}$ on the right-hand side and 
$\{\sigma_{3}\sigma_{4}\sigma_{5}\sigma_{6}\}$ on the left-hand side. $D$ is the total bond dimension.
(b) Two examples of the unitary gates $\hat{U}_{n}$ used in (a). The left one is composed of a single multi-qubit gate with a gate size 
$l=6$, and the right one is composed of two two-qubit gates. 
The number on each red line indicates the bond dimension when the unitary gate is partitioned along this red line. 
(c) A schematic diagram for bipartitioning a unitary gate $\hat{U}_n$.
$D_{n}$ indicates the corresponding bond dimension of $\hat{U}_{n}$.
}
\label{fig:graph-entanglement}
\end{figure*}

Here, we evaluate the upper bound of the entanglement entropy for quantum-circuit states.
Let us assume that the quantum-circuit state $|\Psi\rangle$ is given by $M$-stacked unitary gates 
on the simple product state $|{\mathbf 0}\rangle=|0\rangle^{\otimes L}$ as shown in the left panel of Fig.~\ref{fig:graph-entanglement}(a), i.e., 
\begin{align}
\ket{\Psi}=\hat{U}_{M}\cdots\hat{U}_{1}|{\mathbf 0}\rangle,
\end{align}
where each unitary operator $\hat{U}_{n}$ is a direct product of several multi-qubit gates (or a multi-qubit gate) as shown in Fig.~\ref{fig:graph-entanglement}(b).

To evaluate the von Neumann entropy of the quantum-circuit state, we consider the SVD for the unitary operator $\hat{U}_{n}$ by 
bipartitioning the operator. 
For example, the red line in Fig.~\ref{fig:graph-entanglement}(c) indicates one of several patterns for partitioning the unitary operator 
$\hat{U}_{n}$~\cite{noteunitary}. 
Let us simply write the decomposed unitary operator $\hat{U}_{n}$ as  
\begin{align}
\hat{U}_{n}=\sum^{D_{n}}_{\lambda_{n}=1}d_{\lambda_n}\hat{X}_{n,\lambda_{n}}\hat{Y}_{n,\lambda_{n}},
\label{eq:svd-unitary-op}
\end{align}
where $\hat{X}_{n,\lambda_{n}}$ and $\hat{Y}_{n,\lambda_{n}}$ are non-unitary operators and $d_{\lambda_n}$ is the singular value.
As shown in Fig.~\ref{fig:graph-entanglement}(c), the right (left) indexes of qubits for operator $\hat{X}_{n,\lambda_{n}}$ are denoted as $\{x_{n,i}\}_{1\le i\le I_{n}}$ ($\{x_{n-1,i}\}_{1\le i\le I_{n-1}}$), i.e., 
\begin{align}
\hat{X}_{n,\lambda_n}&=\sum_{\{\sigma_{x_{n-1},i}\}}
\sum_{\{\sigma_{x_n,i}\}} (X_{n,\lambda_n})^{\sigma_{x_n,1}\sigma_{x_n,2}\cdots\sigma_{x_n,I_n}}_{\sigma_{x_{n-1},1}\sigma_{x_{n-1},2}\cdots\sigma_{x_{n-1},I_{n-1}}} \nonumber\\
&\times|\sigma_{x_n,1}\sigma_{x_n,2}\cdots\sigma_{x_n,I_n}\rangle
\langle\sigma_{x_{n-1},1}\sigma_{x_{n-1},2}\cdots\sigma_{x_{n-1},I_{n-1}}|, 
\end{align}
where $I_{n}$ ($I_{n-1}$) specifies the number of right indexes of qubits for $\hat{X}_{n,\lambda_{n}}$ ($\hat{X}_{n-1,\lambda_{n-1}}$), 
implicitly assuming that the number of left indexes of qubits for $\hat{X}_{n,\lambda_{n}}$ is $I_{n-1}$. 
Similarly, the right (left) indexes of qubits for operator $\hat{Y}_{n,\lambda_{n}}$ are denoted as $\{y_{n,i}\}_{1\le i\le L-I_{n}}$ ($\{y_{n-1,i}\}_{1\le i\le L-I_{n-1}}$). 
The bond dimension $D_{n}$ of the unitary gate $\hat{U}_{n}$ is then given as
\begin{align}
D_{n}=&\mathrm{min}\left(\sum_{\bm{x}_{n-1}}\sum_{\bm{x}_{n}}1,\sum_{\bm{y}_{n-1}}\sum_{\bm{y}_{n}}1\right)
\\=&\mathrm{min}\left( 2^{I_{n-1}+I_{n}}, 2^{2L-(I_{n-1}+I_{n})}\right)~.
\end{align}
with $\sum_{\bm{x}_{n}(\bm{y}_{n})}=\sum_{x_{n,1}(y_{n,1})=0,1}\cdots\sum_{x_{n, I_n}(y_{n, L-I_n})=0,1}$.
Note that in counting the bond dimension, one should not take into account the indexes of qubits connecting to the initial qubit state $\ket{0}$ and the indexes of qubits for other gate operators in $\hat{U}_n$ that are intact by the bipartition.
Using these counting rules for the bond dimension, we can calculate the bond dimension of a unitary gate, as shown in 
Fig.~\ref{fig:graph-entanglement}(b).

Using Eq.~\eqref{eq:svd-unitary-op}, we can explicitly write the quantum-circuit state with $M$-stacked unitary gates as follows:
\begin{align}
\ket{\Psi}=&\sum^{D_{1}}_{\lambda_{1}=1}\cdots\sum^{D_{M}}_{\lambda_{M}=1}d_{\lambda_{1}}\cdots d_{\lambda_{M}}
\nonumber\\&\times
(\hat{X}_{M,\lambda_{M}}\cdots\hat{X}_{1,\lambda_{1}})(\hat{Y}_{M,\lambda_{M}}\cdots\hat{Y}_{1,\lambda_{1}})|\mathbf{0}\rangle.
\end{align}
Note that the two sets of vectors
\begin{align}
&\ket{\bm{\lambda}}_{X}=\hat{X}_{M,\lambda_{M}}\cdots\hat{X}_{1,\lambda_{1}}|0\rangle^{\otimes I_{0}} 
\end{align}
and
\begin{align}
&\ket{\bm{\lambda}}_{Y}=\hat{Y}_{M,\lambda_{M}}\cdots\hat{Y}_{1,\lambda_{1}}|0\rangle^{\otimes L-I_{0}}
\end{align}
correspond to the non-orthogonalized basis vectors in the two separated subspaces. 
%Here, the number of left indexes of qubits for $\hat{X}_{n,\lambda_n}$ ($\hat{X}_{n,\lambda_n}$) implies 
%$I_{n-1}$ ($L-I_{n-1}$) and thus 
%$I_0$ is the number of left indexes of qubits for $\hat{X}_{1,\lambda_1}$.
We can orthonormalize these vectors using the Gram-Schmidt method, i.e.,  
\begin{align}
&\ket{\bm{\lambda}}_{X}=\sum^{D_{X}}_{a=1}|u_{a}\rangle\mathcal{X}_{a,\lambda_{M}\cdots\lambda_{1}}
\end{align}
and
\begin{align}
&\ket{\bm{\lambda}}_{Y}=\sum^{D_{Y}}_{b=1}|v_{b}\rangle\mathcal{Y}_{b,\lambda_{M}\cdots\lambda_{1}},
\end{align}
where the total numbers $D_X$ and $D_Y$ of the orthonormalized vectors is given respectively as 
\begin{align}
D_{X}=&\mathrm{rank}({}_{X}\langle\bm{\lambda}|\bm{\lambda}'\rangle_{X})
\le\mathrm{min}(2^{I_{M}}, D_\mathrm{tot})
\end{align}
and
\begin{align}
D_{Y}=&\mathrm{rank}({}_{Y}\langle\bm{\lambda}|\bm{\lambda}'\rangle_{Y})
\le\mathrm{min}(2^{L-I_{M}}, D_\mathrm{tot})
\end{align}
with 
\begin{align}
D_\mathrm{tot}=\prod^{M}_{n=1}D_{n}.
\label{eq:dtotal}
\end{align}
The quantum-circuit state $\ket{\Psi}$ is then written as
\begin{align}
|\Psi\rangle=&\sum^{D_{X}}_{a=1}\sum^{D_{Y}}_{b=1}W_{ab}|u_{a}\rangle\otimes|v_{b}\rangle
\end{align}
with
\begin{align}
W_{ab}=&\sum_{\lambda_{1},\cdots,\lambda_{M}}d_{\lambda_{1}}\cdots d_{\lambda_{M}}
\mathcal{X}_{a,\lambda_{M}\cdots\lambda_{1}}\mathcal{Y}_{b,\lambda_{M}\cdots\lambda_{1}}.
\end{align}

Therefore, performing the SVD for $W_{ab}=\sum^{D}_{\lambda=1}w_{\lambda}\tilde{\mathcal{X}}_{\lambda,a}\tilde{\mathcal{Y}}_{\lambda,b}$ with $D=\mathrm{min}(D_{X},D_{Y})$,
the quantum-circuit state finally reads
\begin{align}
|\Psi\rangle=\sum^{D}_{\lambda=1}w_{\lambda}|\tilde{u}_{\lambda}\rangle\otimes|\tilde{v}_{\lambda}\rangle,
\label{eq:Psi_w}
\end{align}
where $|\tilde{u}_{\lambda}\rangle=\sum^{D_X}_{a=1}|u_a\rangle\tilde{\mathcal{X}}_{\lambda,a}$ and
$|\tilde{v}_{\lambda}\rangle=\sum^{D_Y}_{b=1}|v_b\rangle\tilde{\mathcal{Y}}_{\lambda,b}$.
Thus, the upper bound of the von Neumann entropy $S$ of the quantum-circuit state is given by the logarithm of the bond dimension $D$, i.e.,
\begin{align}
S=&-\sum^{D}_{\lambda=1}w^2_{\lambda}\ln{w^{2}_{\lambda}}\le \ln{D}
\end{align}
with
\begin{align}
D\le&\min(2^{\mathrm{min}(I_{M}, L-I_{M})},D_\mathrm{tot}),
\end{align}
where the normalization condition $\sum^{D}_{\lambda=1}w_{\lambda}^2=1$ is implied. 
In particular, when no path is found with the bond dimension $D_\mathrm{tot}<2^{\mathrm{min}(I_{M}, L-I_{M})}$ 
for the 1D bipartition $\{\sigma_{1}\cdots\sigma_{I_{M}}|\sigma_{I_{M}+1}\cdots\sigma_{L}\}$, 
we consider that the quantum circuit state follows the volume law in entanglement entropy.

\begin{figure}[htbp]
\includegraphics[width=.9\columnwidth]{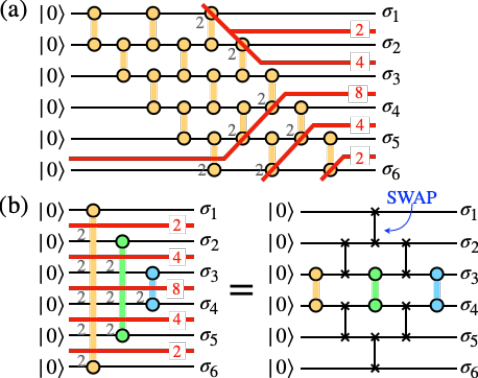}
\caption{
Two examples of quantum-circuit ansatze satisfying the volume law.
(a) A three-layer sequential-type quantum circuit composed of two-qubit gates.
(b) A quantum circuit consisting of the minimum number of two-qubit gates, i.e., three two-qubit gates (left panel),
which is equivalent to a 
diamond-shaped quantum circuit with swap gates (right panel).
The red lines indicate several paths for the 1D bipartition and
the number on each red line denotes the corresponding bond dimension $D_\mathrm{tot}$ 
(indicated by red) and $\{D_{n}\}_{1\le n\le M}$ (indicated by black). 
}
\label{fig:graph-entanglement-example}
\end{figure}

As discussed in Sec.~\ref{sec:diamond}, a quantum-circuit state satisfying the volume law can represent a highly-entangled quantum state, often appearing in a time-evolved state for non-equilibrium dynamics.
Hence, it is instructive to illustrate several quantum-circuit structures that satisfy the volume law.
Figure~\ref{fig:graph-entanglement-example} shows two quantum-circuit states satisfying the volume law.
The simplest example is a sequential-type quantum circuit composed of two-qubit gates with multiple layers shown 
in Fig.~\ref{fig:graph-entanglement-example}(a).
If we calculate the bond dimension for the 1D bipartition shown by the red line in Fig.~\ref{fig:graph-entanglement-example}(a), 
the bond dimension $D_\mathrm{tot}$ in Eq.~(\ref{eq:dtotal}) 
satisfies $D_\mathrm{tot}=2^{\mathrm{min}(I_{M}, L-I_{M})}$. 
On the other hand, if we consider a quantum circuit that is one layer less than the quantum circuit 
in Fig.~\ref{fig:graph-entanglement-example}(a), we can easily confirm that there exists a path across the two layers that 
results in a loss of the volume law.
Thus, the sequential-type quantum circuit composed of two-qubit gates that satisfies the volume law needs $m=\lfloor\frac{L}{2}\rfloor$ layers, e.g., $3$ layers for $L=6$ and $5$ layers for $L=11$.
In other words, the sequential-type quantum circuit containing the diamond-shaped quantum circuit satisfies the volume law [see Fig.~\ref{fig:diamond-decomposition}(c)].

In addition to the above example, the quantum-circuit structure shown in the left panel of Fig.~\ref{fig:graph-entanglement-example}(b) 
also satisfies the volume law.
One can easily show that this quantum circuit is transformed into a diamond-shaped quantum circuit with swap gates 
as given in the right panel of Fig.~\ref{fig:graph-entanglement-example}(b).
In particular, if we replace all these two-qubit gates in the quantum circuit with the products of Hadamard and controlled-NOT gates, the resulting quantum-circuit state is maximally entangled.
However, a quantum circuit satisfying the volume law is not a sufficient condition for representing a highly-entangled quantum state but only a necessary condition.
Since the quantum circuit in the left panel of Fig.~\ref{fig:graph-entanglement-example}(b) comprises three independent subspaces, i.e., $\{\sigma_{1}\sigma_{6}\}$, $\{\sigma_{2}\sigma_{5}\}$ and $\{\sigma_{3}\sigma_{4}\}$, it is clear that such a quantum-circuit state 
can not represent a quantum state with physically relevant correlations among these subspaces.

\section{\label{app:optimize_each_timestep} Optimized Quantum Circuits for Exact Quantum Dynamics}

As shown in Figs.~\ref{fig:evaluation-diamond}(c) and \ref{fig:evaluation-diamond}(d), we find that the diamond-shaped quantum circuit satisfying the volume law can exactly represent the time-evolved states for relatively long-time dynamics of the transverse field Ising model without the longitudinal field (i.e., $h=0$).  
This appendix confirms that such behavior is not observed for the multi-layer sequential-type quantum circuits composed of two-qubit gates having the same degrees of freedom as the diamond-shaped quantum circuits and for the single-layer sequential-type quantum circuits composed of multi-qubits ($l>2$) gates having more internal degrees of freedom than the diamond-shaped quantum circuits but without showing the volume law. 
Note that the quantum circuits $\hat{C}_t$ studied here are optimized by minimizing the cost function $F=\left\| |\Psi(t)\rangle - \hat{C}_t |\Phi\rangle \right\|^2$, directly referring to the exact time-evolved state $|\Psi(t)\rangle$, instead of the cost function given in 
Eq.~(\ref{eq:i}) and equivalently in Eq.~(\ref{eq:Fcost}).

\begin{figure}[htbp]
\includegraphics[width=1.0\columnwidth]{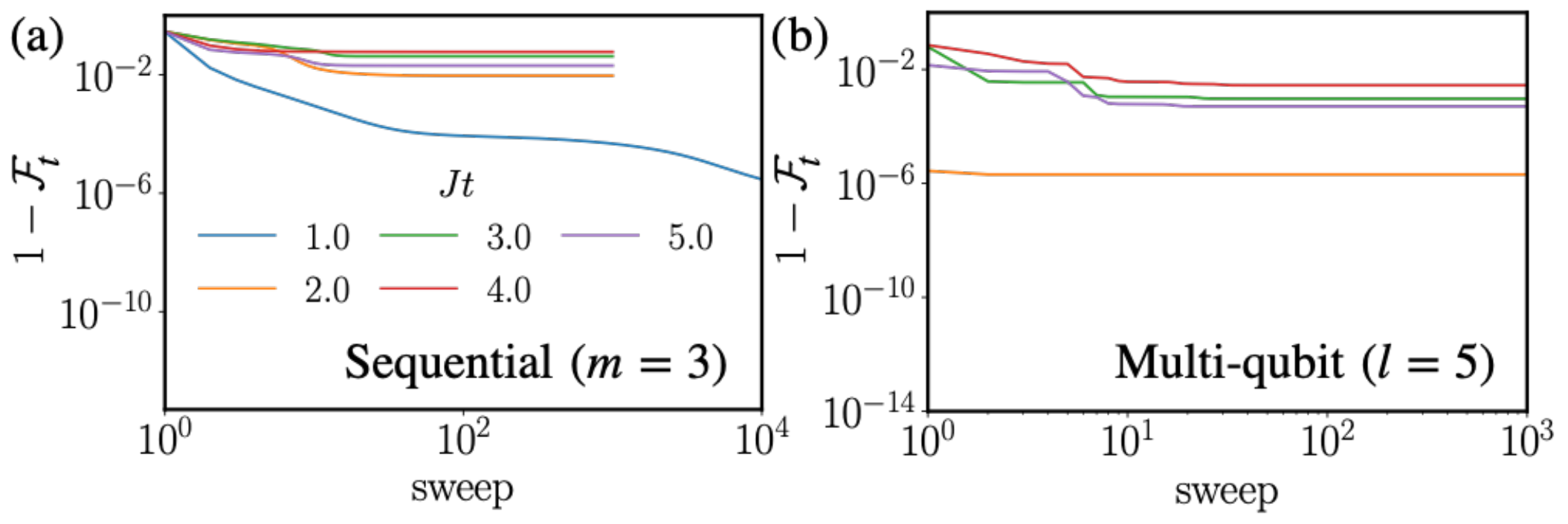}
\caption{
The infidelity $1-\mathcal{F}_t$ of the quantum-circuit states with (a) the three-layer sequential-type quantum circuit composed of 
two-qubit gates and (b) the single-layer sequential-type quantum circuit composed of multi-qubit gates (the gate size $l=5$) 
at different times $Jt=1,2, \dots,5$ in quantum dynamics 
for the $L=11$ quantum Ising chain with the model parameters $g=1.4$ and $h=0$. 
The hyperparameters are set to be $w_\mathrm{max}=10^{4}$, $w_\mathrm{min}=10^{3}$, $\epsilon_{a}=10^{-14}$, 
and $\epsilon_{r}=10^{-4}$. 
These model parameters and hyperparameters are the same as those in Fig.~\ref{fig:evaluation-diamond}(a).
Note that the result for $Jt=1$ in (b) is smaller than $10^{-14}$.
}
\label{fig:qce-fixed-time-l11-l16}
\end{figure}

Figure~\ref{fig:qce-fixed-time-l11-l16}(a) shows the infidelity $1-{\mathcal{F}}_t$ for the transverse-field Ising model with the transverse field $g=1.4$ for $L=11$ as a function of the number of sweep iteration updates for the optimization of the circuit state $|\Phi(t)\rangle=\hat{C}_t|\Phi\rangle$ with the three-layer ($m=3$) sequential-type quantum circuit composed of two-qubit gates, 
which has the same number of two-qubit gates, $n_{g}=30$, of the diamond-shaped quantum circuit.
At $Jt=1$, when the growth of entanglement is not large, the sequential-type quantum circuit can well represent the exact time-evolved quantum state $|\Psi(t)\rangle$ with its infidelity as small as $10^{-6}$. However, for $Jt \geq 2$, when the growth of entanglement becomes pronounced, the quantum-circuit ansatz based on the sequential-type quantum circuit has large infidelity about $10^{-2}$, which causes the rather large error accumulated in the long-time dynamics. 
As shown in Fig.~\ref{fig:qce-fixed-time-l11-l16}(b), similar behavior is observed in the result for the single-layer sequential-type quantum circuit composed of $5$-qubit gates ($l=5$), corresponding to an MPS with the bond dimension 16.
This is understood simply because in order for the MPS representation to exactly represent an arbitrary quantum state, a bond dimension larger than $2^{\lfloor L/2 \rfloor}$ is required.

On the other hand, as discussed in Sec.~\ref{sec:diamond}, the quantum-circuit state based on the diamond-shaped quantum circuit follows the volume law. Moreover, the internal degrees of freedom in the diamond-shaped quantum circuit are smaller than those in the single-layer sequential-type quantum circuit composed of multi-qubit gates representing exactly an arbitrary quantum state. 
Therefore, the diamond-shaped quantum circuit or a quantum circuit with a similar structure [e.g., see Fig.~\ref{fig:diamond-decomposition}(c)] can potentially be a powerful component of quantum circuits to achieve quantum advantage, at least in quantum many-body physics and especially non-equilibrium quantum many-body dynamics.

% \bibliography{qce-2}
\bibliography{qce,notes}

\begin{thebibliography}{37}
\bibitem [{not()}]{noteunitary}%
  \BibitemOpen
  \href@noop {} {}\bibinfo {note} {More precisely, the matrix representing the
  unitary operator $\hat{U}_n$ is singular value decomposed and the resulting
  matrices are used to construct the singular value decomposed $\hat{U}_n$ in
  Eq.~(\ref{eq:svd-unitary-op}).}\BibitemShut {Stop}%
\end{thebibliography}

\end{document}